\documentclass[twocolumn]{aastex631}
\usepackage{verbatim}
\usepackage{graphicx}
\usepackage{upgreek}
\usepackage{natbib}
\usepackage{xcolor}
\usepackage[version=4]{mhchem}
 \usepackage{amsmath}
\definecolor{darkred}{rgb}{0.55, 0.0, 0.0} 
\definecolor{royalblue}{rgb}{0.25, 0.41, 0.88}
\shorttitle{ }
\shortauthors{A. Gressier, et al.}
\defcitealias{May_2023}{May \& MacDonald et al. 2023}
\defcitealias{Moran_2023}{Moran \& Stevenson et al. 2023}
\accepted{for publication in the Astronomical Journal, September 16, 2025.}

\begin{document}

\title{JWST-TST DREAMS: Sulfur dioxide in the atmosphere of the Neptune-mass planet HAT-P-26\,b from NIRSpec G395H transmission spectroscopy}


\author[0000-0003-0854-3002]{Am\'{e}lie Gressier}
\affiliation{Space Telescope Science Institute, 3700 San Martin Drive, Baltimore, MD 21218, USA}

\author[0000-0003-1240-6844]{Natasha E. Batalha}
\affiliation{NASA Ames Research Center, MS 245-3, Moffett Field, CA 94035, USA}

\author[0000-0002-0413-3308]{Nicholas Wogan}
\affiliation{NASA Ames Research Center, MS 245-3, Moffett Field, CA 94035, USA}

\author[0000-0001-8703-7751]{Lili Alderson}
\affiliation{Department of Astronomy, Cornell University, 122 Sciences Drive, Ithaca, NY 14853, USA}

\author[0009-0003-1341-3736]{Dominic Doud}
\affiliation{NASA Ames Research Center, MS 245-3, Moffett Field, CA 94035, USA}


\author[0000-0001-9513-1449]{N\'estor Espinoza}
\affiliation{Space Telescope Science Institute, 3700 San Martin Drive, Baltimore, MD 21218, USA}
\affiliation{William H. Miller III Department of Physics and Astronomy, Johns Hopkins University, Baltimore, MD 21218, USA}

\author[0000-0003-4816-3469]{Ryan J. MacDonald}
\affiliation{Department of Astronomy, University of Michigan, 1085 S. University Ave., Ann Arbor, MI 48109, USA}
\affiliation{School of Physics and Astronomy, University of St Andrews, North Haugh, St Andrews, KY16 9SS, UK}

\author[0000-0003-4328-3867]{Hannah R. Wakeford}
\affiliation{University of Bristol, HH Wills Physics Laboratory, Tyndall Avenue, Bristol, UK}

\author[0000-0003-3305-6281]{Jeff A. Valenti}
\affiliation{Space Telescope Science Institute, 3700 San Martin Drive, Baltimore, MD 21218, USA}

\author[0000-0002-8507-1304]{Nikole K. Lewis}
\affiliation{Department of Astronomy and Carl Sagan Institute, Cornell University, 122 Sciences Drive, Ithaca, NY 14853, USA}

\author{Sara Seager}
\affiliation{Department of Physics and Kavli Institute for Astrophysics and Space Science, Massachusetts Institute of Technology, 77 Massachusetts Ave, Cambridge, MA, 02139, USA}
\affiliation{Department of Earth, Atmospheric and Planetary Sciences, Massachusetts Institute of Technology, 77 Massachusetts Ave, Cambridge, MA, 02139, USA} 
\affiliation{Department of Aeronautics and Astronautics, Massachusetts Institute of Technology, 77 Massachusetts Avenue, Cambridge, MA 02139, USA}

\author[0000-0002-7352-7941]{Kevin B. Stevenson}
\affiliation{Johns Hopkins University Applied Physics Laboratory, Laurel, MD 20723, USA}

\author[0000-0002-0832-710X]{Natalie H. Allen}
\affiliation{William H. Miller III Department of Physics and Astronomy, Johns Hopkins University, Baltimore, MD 21218, USA}

\author[0000-0003-4835-0619]{Caleb I. Ca\~nas}
\affiliation{Exoplanets and Stellar Astrophysics Laboratory (Code 667), NASA Goddard Space Flight Center, Greenbelt, MD 20771, USA}

\author[0000-0002-8211-6538]{Ryan C. Challener}
\affiliation{Department of Astronomy and Carl Sagan Institute, Cornell University, 122 Sciences Drive, Ithaca, NY 14853, USA}

\author[0000-0002-5322-2315]{Ana Glidden}
\affiliation{Department of Earth, Atmospheric and Planetary Sciences, Massachusetts Institute of Technology, Cambridge, MA 02139, USA}

\author[0000-0001-5732-8531]{Jingcheng Huang}
\affiliation{Department of Earth, Atmospheric and Planetary Sciences, Massachusetts Institute of Technology, Cambridge, MA 02139, USA}

\author[0000-0003-0525-9647]{Zifan Lin}
\affiliation{Department of Earth, Atmospheric and Planetary Sciences, Massachusetts Institute of Technology, Cambridge, MA 02139, USA}

\author[0000-0002-2457-272X]{Dana R. Louie}
\affiliation{Catholic University of America, Department of Physics, Washington, DC, 20064, USA}
\affiliation{Exoplanets and Stellar Astrophysics Laboratory (Code 667), NASA Goddard Space Flight Center, Greenbelt, MD 20771, USA}
\affiliation{Center for Research and Exploration in Space Science and Technology II, NASA/GSFC, Greenbelt, MD 20771, USA}

\author[0000-0002-2643-6836]{Cathal Maguire}
\affiliation{University of Bristol, HH Wills Physics Laboratory, Tyndall Avenue, Bristol, UK}

\author[0000-0003-0814-7923]{Elijah Mullens}
\affiliation{Department of Astronomy and Carl Sagan Institute, Cornell University, 122 Sciences Drive, Ithaca, NY 14853, USA}

\author[0000-0002-7352-7941]{Kristin Sotzen}
\affiliation{Johns Hopkins University Applied Physics Laboratory, Laurel, MD 20723, USA}

\author[0000-0002-2643-6836]{Daniel Valentine}
\affiliation{University of Bristol, HH Wills Physics Laboratory, Tyndall Avenue, Bristol, UK}

\author{Mark Clampin}
\affiliation{Science Mission Directorate, Mary W. Jackson NASA Headquarters 300 E Street SW, Washington, DC 20546, USA}

\author{Laurent Pueyo}
\affiliation{Space Telescope Science Institute, 3700 San Martin Drive, Baltimore, MD 21218, USA}

\author[0000-0001-7827-7825]{Roeland P. van der Marel}
\affiliation{Space Telescope Science Institute, 3700 San Martin Drive, Baltimore, MD 21218, USA}
\affiliation{Center for Astrophysical Sciences, The William H. Miller III Department of Physics \& Astronomy, Johns Hopkins University, Baltimore, MD 21218, USA}

\author{C. Matt Mountain}
\affiliation{Association of Universities for Research in Astronomy, 1331 Pennsylvania Avenue NW Suite 1475, Washington, DC 20004, USA}

\begin{abstract}
We present the James Webb Space Telescope (JWST) transmission spectrum 
of the exoplanet HAT-P-26\,b (18.6\,M$_\oplus$, 6.33\,R$_\oplus$), based on a single transit observed with the JWST NIRSpec G395H grating. We detect water vapor ($\ln \mathcal{B}$=4.1), carbon dioxide ($\ln \mathcal{B}$=85.6), and sulfur dioxide ($\ln \mathcal{B}$=13.5) with high confidence, along with marginal indications for hydrogen sulfide and carbon monoxide ($\ln \mathcal{B}<$0.5). 
The detection of SO$_2$ in a warm super-Neptune sized exoplanet (R$_{\rm P}\sim$6\,R$_{\oplus}$) bridges the gap between previous detections in hot Jupiters and sub-Neptunes, highlighting the role of disequilibrium photochemistry across a broad range of exoplanet atmospheres, including those cooler than 1000\,K. Our precise measurements of carbon, oxygen, and sulfur indicate an atmospheric metallicity of $\sim$10$\times$ solar and a sub-solar C/O ratio. 
Retrieved molecular abundances are consistent within 2$\sigma$ with predictions from self-consistent models including photochemistry. The elevated CO$_2$ abundance and possible H$_2$S signal may also reflect sensitivities to the thermal structure, cloud properties, or additional disequilibrium processes such as vertical mixing. We compare the SO$_2$ abundance in HAT-P-26 b with that of ten other JWST-observed giant exoplanets, and find a correlation with atmospheric metallicity. The trend is consistent with the prediction from \citet{Crossfield_2023}, showing a steep rise in SO$_2$ abundance at low metallicities, and a more gradual increase beyond 30$\times$ solar. 
This work is part of a series of studies by our JWST Telescope Scientist Team (JWST-TST), in which we use Guaranteed Time Observations to perform Deep Reconnaissance of Exoplanet Atmospheres through Multi-instrument Spectroscopy (DREAMS).
\end{abstract}

\section{Introduction} \label{sec:intro}
In just over three decades, exoplanetary science has progressed from the discovery of gas giants to the characterization of smaller, temperate exoplanets. 
While small planets (R$_{\rm P}\sim$6R$_{\oplus}$) are prevalent in our galaxy \citep{Fulton_2017, Fulton_2018, Van_Eylen_2018}, their formation remains a key unresolved question. One way to investigate this is through atmospheric characterization of transiting exoplanets. The atmospheric composition of exoplanets, traced through the abundances of key molecules like carbon and oxygen, provides insight into planet formation in the protoplanetary disk \citep{Oberg_2011, Madhusudhan_2014, Mordasini_2016}. This is where the James Webb Space Telescope (JWST) is particularly revolutionary. In-depth spectroscopic studies of sub-Neptune to Neptune-mass planets are now feasible with JWST, enabling unprecedented constraints on atmospheric composition, structure, metallicity, C/O ratio, and cloud properties \citep[e.g.][]{Madhusudhan_2023, holmberg_2024, benneke2024jwst, Beatty_2024, davenport_2025, madhusudhan_2025, hu_2025}. While these efforts have primarily focused on sub-Neptunes ($<$4\,R$\oplus$, $<$10\,M$\oplus$), observations of Neptune-mass planets remain sparse. To date, only a few such planets have been characterized with JWST, including GJ\,3470\,b in transmission \citep{Beatty_2024} and GJ\,436\,b in emission \citep{Mukherjee_2025}, leaving a significant gap in our understanding of this mass regime. The present study of HAT-P-26\,b marks only the second JWST atmospheric characterization in transmission of a Neptune-mass planet, and the first for a super-Neptune in this radius and temperature range, providing a critical new benchmark for atmospheric studies in this regime. While LTT\,9779\,b has also been observed with JWST \citep{Radica_2024, Coulombe_2025}, it lies in the Neptune desert and has a mass above $\sim$30\,M$_\oplus$, making it substantially more massive, denser, and hotter than typical Neptune-mass planets.

The Hubble Space Telescope (HST) played a central role in characterizing the atmosphere of gas giant exoplanets. The Wide Field Camera 3 (WFC3) enabled detections of molecular species such as water, while the Space Telescope Imaging Spectrograph (STIS) provided key constraints in the optical. These early observations, alongside contributions from Spitzer, laid essential groundwork for the more detailed characterization now possible with JWST’s enhanced spectral coverage and precision. 
However, degeneracies between cloud or haze coverage and atmospheric metallicity often limited the ability to constrain molecular abundances. With JWST, these constraints are now possible, as illustrated in the review by \citet{espinoza_2025}. 

Beyond providing access to precise carbon and oxygen abundances, JWST has expanded the usual molecular inventory to a new chemical regime: sulfur chemistry. Sulfur-bearing species, in particular SO$_2$ and H$_2$S, have been detected and their abundances constrained using JWST in transmission spectroscopy. These signatures, absent in brown dwarfs at similar temperatures, point to unique photochemical processes in highly irradiated exoplanets. Observations with NIRSpec and NIRCam in the near-infrared \citep[e.g.,][]{Rustamkulov_2023, Alderson_2023, Sing_2024, Beatty_2024, fu_2024} have revealed these species in the hot Jupiters WASP-39\,b, WASP-107\,b, and HD\,189733\,b, as well as in the Neptune GJ 3470\,b. In particular, the $\sim$4.0\,$\upmu$m sulfur dioxide absorption band, covered by the G395H grating, has been identified as a key indicator of sulfur chemistry. MIRI LRS observations, above 5\,$\upmu$m, have also enabled the detection of SO$_2$ for WASP-39\,b \citep{Powell_2024} and WASP-107\,b \citep{Dyrek_2023}. Alongside these detections, theoretical studies have shown that H$_2$S and SO$_2$ are the most likely sulfur species to be detectable with JWST \citep{Polman_2023}, that SO$_2$ in WASP-39\,b is robustly explained as a photochemical product of H$_2$S destruction \citep{Tsai_2023}, and that its abundance provides a sensitive probe of metallicity and planet formation pathways \citep{Crossfield_2023}. More recent work further highlights multiple distinct chemical pathways to SO$_2$ formation, driven by the stellar UV flux at different wavelengths \citep{de_Gruijter_2025}, emphasizing the role of photochemistry and stellar environment in shaping the observed sulfur chemistry.

HAT-P-26\,b, with a mass of 18.6 M$_\oplus$ and a radius of 6.33 R$_\oplus$, is one of the prototypical low-mass giant planets showing spectroscopic features in its HST transmission spectrum \citep{Hartman_2011, Wakeford_2017}. 
Its extended atmosphere, resulting from a warm temperature (T$_{\rm eq}\,\sim$\,1000\,K) and low gravity, made it one of the most studied low-mass exoplanets before the launch of JWST \citep[e.g.][]{Stevenson_2016b, Wakeford_2017, Tsiaras_2018, MacDonald_2019, ramos_rosado_2025}. Its transmission spectrum, compiled from HST, Magellan, and Spitzer observations, covers a discontinuous wavelength range from 0.3 to 5\,$\upmu$m. These data enabled the detection of water and contributed to the first constraints on the planet’s atmospheric metallicity, initially estimated at $4.8^{+21.5}_{-4.0} \times$ solar by \citet{Wakeford_2017} and later refined to $18.1^{+25.9}_{-11.3} \times$ solar in a reanalysis by \citet{MacDonald_2019}. In this work, we present the NIRSpec G395H transmission spectrum of HAT-P-26\,b, which is expected to exhibit strong spectral features of water and carbon dioxide based on equilibrium chemistry calculations and constraints from previous studies \citep{Stevenson_2016b, Wakeford_2017, MacDonald_2019}. The observations are described in Section\,\ref{sec:observation}, while the steps taken to obtain the transmission spectrum are detailed in Section\,\ref{sec:DataAnalysis}. The atmospheric interpretation, including forward modeling and retrieval techniques, is presented in Section\,\ref{sec:interpretation}. The chemical composition derived from the near-infrared spectrum provides surprising new insights, which are discussed in Section\,\ref{sec:discussion}, particularly in relation to other JWST analyses.

This paper is part of a series by the JWST Telescope Scientist Team (JWST TST), \footnote{\url{https://www.stsci.edu/~marel/jwsttelsciteam.html}} which uses
Guaranteed Time Observer (GTO) time awarded by NASA in 2003 (PI: M. Mountain) for studies in three different subject areas: (a) Transiting Exoplanet Spectroscopy (lead: N. Lewis); (b) Exoplanet and Debris Disk Coronagraphic Imaging (lead: M. Perrin); and (c) Local Group Proper Motion Science (lead: R. van der Marel). The present paper is part of our work on Transiting Exoplanet Spectroscopy, which focuses on detailed exploration of three transiting exoplanets representative of key exoplanet classes: Hot Jupiters (WASP-17b, GTO~1353), Warm Neptunes (HAT-P-26b, GTO~1312), and Temperate Terrestrials (TRAPPIST-1e, GTO~1331). Complementary analyses of the NIRISS SOSS (MacDonald et al., in prep.) and full transmission spectrum including NIRISS SOSS, NIRSpec G395H, and MIRI LRS (Alderson et al., in prep.) will be presented in forthcoming papers.


\section{Observations} \label{sec:observation}
We used the NIRSpec Bright Object Time Series (BOTS) mode with the G395H grating and F290LP long-pass filter to observe a transit of HAT-P-26\,b on June 15, 2023, as part of program 1312 (PI: N. Lewis). The observation spanned 7 hours and 56 minutes, including the 2-hour and 30-minute transit duration, and a sufficient baseline to account for systematics and uncertainties in the mid transit time. The exposure consisted of 481 integrations, each with 48 single-frame (NRSRAPID) groups up the ramp. We used the SUB2048 subarray, which samples 2048 by 32 pixels on each detector that the spectrum is dispersed across (NRS1, NRS2). 
\begin{figure*}
    \centering    
    \includegraphics[width=\textwidth]{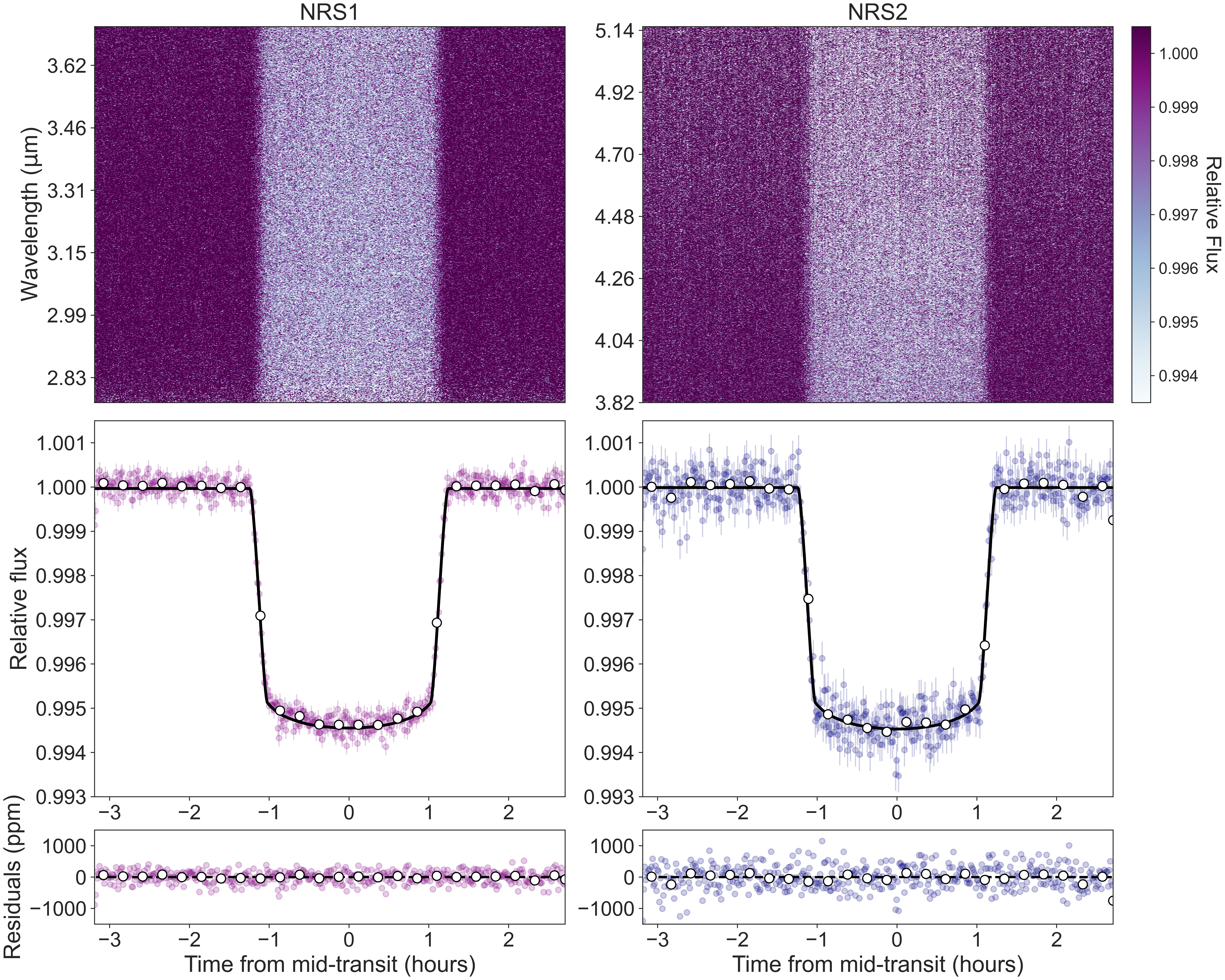}
    \caption{Spectroscopic and broadband light curves for the JWST NIRSpec G395H transit observation of HAT-P-26\,b. Top: Raw spectroscopic light curves for NRS1 and NRS2 from the \texttt{transitspectroscopy} reduction. Middle: De-trended broadband light curves with the best-fit transit model shown in black. Bottom: Residuals between the broadband light curves and the best-fit model. Data points at full temporal resolution are shown in purple (NRS1) and blue (NRS2) with their associated uncertainties. White data points represent time-binned data for clarity.}
    \label{fig:wlc_transitspectroscopy}
\end{figure*}
\section{JWST NIRSpec G395H data analysis} \label{sec:DataAnalysis}
We performed two different reductions of the data for HAT-P-26\,b from program 1312. In the following sections, we outline the different steps for each reduction.

\subsection{\texttt{transitspectroscopy} Data Reduction} \label{sec:2.1}
The first reduction and data analysis were performed using the open-source \texttt{Python} package \texttt{transitspectroscopy} \citep{espinoza_2022}. Specifically, we used Stage 1 of \texttt{transitspectroscopy}, which serves as a wrapper for the \texttt{jwst} calibration pipeline, version 1.14.0. Our analysis starts with the `uncal' FITS file downloaded from the Mikulski Archive for Space Telescopes (MAST) for both NRS1 and NRS2 detectors. In Stage 1, default data calibration corrections were applied. The jump step is replaced by a custom function that detects jumps by calculating differences between consecutive groups for each pixel. A median-filtered difference is used to isolate noise, and significant outliers are flagged as jumps in the `groupdq' extension. This process is repeated for each group from the first to the second-to-last. Finally, we fit the ramps using the default function from the \texttt{jwst} pipeline.

After applying the initial calibration steps and obtaining the slopes (equivalent to the `rateints'), we proceed to Stage 2, where the raw integration images are converted into time series light curves. This process is applied consistently to both detectors. For each frame, pixels identified as NaN are replaced with zeros to preserve the total variance. We then compute the median frame from these integrations to establish a baseline for background estimation. The trace of the spectrum for each detector is determined using the \texttt{trace\_spectrum} function from the \texttt{transitspectroscopy} package, followed by smoothing with a spline function. This method uses cross-correlation with a Gaussian profile, tracing from columns `xstart' to `xend' and identifying the maximum of the cross-correlation function and the profile across columns. The trace was extracted between 500 and 2042 x-pixels for NRS1 and between 5 and 2043 x-pixels for NRS2 with a 3-pixels radius extraction.

To assess the background, we start by masking pixels in the median rate frame to differentiate between those within and outside the spectral trace. Pixels inside the trace are set to NaN in the out-of-trace mask, while those outside the trace are retained for background estimation. The background is calculated as the median of the remaining out-of-trace pixels and is subsequently subtracted from each integration frame in the time series observation.

For 1/f noise removal \citep{Jakobsen_2022, Birkmann_2022} -- noise arising from the electronics of the readout pattern, which appears as column striping in the subarray image -- we first subtract the median frame from each individual integration frame. We then iterate through the spectral trace, masking out pixels within a defined inner radius (3 pixels) around the trace center and those beyond an outer radius (8 pixels) from the trace. The 1/f noise component is estimated as the median of the remaining pixels in each column, and this estimated noise is subtracted from each individual frame to correct the data. We use a simple box extraction method with a 3-pixel radius to obtain the stellar flux for both NRS1 and NRS2 and outliers exceeding a 5$\sigma$ threshold are replaced with a 1D median spectrum. We then generate pixel light curves for each column of the detectors, as well as broadband light curves by integrating the flux over all pixels in each detector.\\

We fit the light curves from both detectors using the Python package \texttt{juliet} \citep{Espinoza_2019}, which uses the \texttt{batman} package \citep{Kreidberg_2015} for modeling transits and eclipses. The spectral light curves are binned prior to fitting with a 10-pixel bin. Orbital parameters—period ($P$), mid-transit time ($t_0$), scaled semi-major axis ($a/R_{\star}$), and impact parameter—are fixed to the refined values presented in Table\,\ref{table:setup_fit}. These values are derived from a joint radial velocity (RV) and transit fit (Valenti et al., in prep.) using all available data on HAT-P-26\,b. We fit the planet-radius ratio ($R_{P}/R_{\star}$) with uniform priors ranging from 0 to 0.2. Limb-darkening coefficients \(u_1\) and \(u_2\), for the quadratic-law, are fitted freely using a uniform prior between -3 and 3 \citep{coulombe_2024b}. The fit also includes a mean out-of-transit flux factor ($m_{\text{flux}}$), modeled with a normal distribution (mean 0, standard deviation 0.1), and a jitter parameter for white noise ($\sigma_w$) fitted using a log-uniform distribution between 10 and 10000 in ppm. Systematic trends are addressed using a Gaussian Process (GP) with a Matérn 3/2 kernel via the \texttt{celerite} package \citep{Foreman_Mackey_2017}, with log-uniform priors for the GP amplitude ($\text{GP}_{\sigma}$) and length-scale ($\text{GP}_{\rho}$) between \(10^{-5}\) and \(10^{3}\). The extracted 10 pixels bin transit depths and corresponding errors represent the median and variance of the posterior distribution of $R_{P}/R_{\star}$ squared for each wavelength bin. 

We use a similar setup to fit the broadband light curves from NRS1 and NRS2 separately, ensuring that the detrending models are independent. The light curve fitting setup and posterior parameters for NRS1 and NRS2 broadband light curves are detailed in Table\,\ref{table:setup_fit}. Figure\,\ref{fig:wlc_transitspectroscopy} shows the spectroscopic and broadband light curves from the \texttt{transitspectroscopy} reduction used to extract the transmission spectrum of HAT-P-26\,b.

\begin{deluxetable*}{lCCC}
\label{table:setup_fit}
\tablecaption{Light curve fitting setup and posterior parameters for NRS1 and NRS2 broadband light curves from \texttt{transitspectroscopy} data reduction. }
\tablehead{ 
    \colhead{Parameters} & \colhead{Priors} & \colhead{NRS1} & \colhead{NRS2} 
}
\startdata
\hline
\multicolumn{4}{l}{\textbf{Planetary Parameters fixed from global fit}} \\
Period (days), $P$ & \textnormal{fixed} & \multicolumn{2}{c}{4.2344923} \\
Mid-transit time (BJD$_{\rm TDB}$), $t_0$ & \textnormal{fixed} & \multicolumn{2}{c}{2460110.810782}  \\
Impact parameter, $b$ & \textnormal{fixed} & \multicolumn{2}{c}{0.481}  \\
Scaled semi major axis $a/R_{\star}$ & \textnormal{fixed} & \multicolumn{2}{c}{12.52}  \\
Eccentricity $e$ & \textnormal{fixed} & \multicolumn{2}{c}{0}  \\
\hline
\multicolumn{4}{l}{\textbf{Posterior parameters for JWST NIRSpec G395H transit fit}} \\ 
Planet-to-star radius ratio, $R_{P}/R_{\star}$ & $\mathcal{U}(0, 0.2)$ &0.07207$^{+0.00025}_{-0.00027}$ &0.07227$^{+0.00034}_{-0.00031}$ \\ 
Limb-darkening coefficient, $u1$ & $\mathcal{U}(-3, 3)$ & {0.1544 $^{+0.0676}_{-0.0678}$}& {0.1687 $^{+0.1230}_{-0.1293}$} \\
Limb-darkening coefficient, $u2$ & $\mathcal{U}(-3, 3)$ & {0.0619 $^{+0.0869}_{-0.0838}$} & {0.0548 $^{+0.1646}_{-0.1590}$}\\
Out-of-transit flux $m_{\text{flux}}$ & $\mathcal{N}(0.0, 0.1)$ & 0.000051$^{+0.0019}_{-0.0018}$  & 0.000010$^{+0.00026}_{-0.00021}$\\
White Noise (ppm) $\sigma_w$ & $\mathcal{LU}(10, 10000)$ & 171.03 $^{+7.39}_{-7.11}$ & 351.71 $^{+13.00}_{-12.73}$ \\
GP amplitude $\text{GP}_{\sigma}$  & $\mathcal{LU}(10^{-5}, 1000)$ & 0.0013$^{+0.0058}_{-0.0011}$& 0.00012$^{+0.0011}_{-0.0001}$ \\
GP length scale $\text{GP}_{\rho}$ & $\mathcal{LU}(10^{-5}, 1000)$ & 34.45$^{+302.21}_{-31.68}$ & 43.09$^{+360.58}_{-42.15}$ \\
\hline
\multicolumn{4}{l}{\textbf{Statistical results for JWST NIRSpec G395H transit fit}} \\
ln$\mathcal{Z}$ GP + transit fit & - & 3398  & 3092 \\
ln$\mathcal{Z}$ transit only & - & 3386  & 3089 \\
ln$\mathcal{Z}$ GP only & - & 3296  & 3012 \\
ln$\mathcal{Z}$ linear ramp + transit fit & - &  3390 & 3079 \\
\hline
\enddata
\tablecomments{The fit is performed using the Python package \texttt{juliet} \citep{Espinoza_2019} \footnote{\url{https://github.com/nespinoza/juliet}}
}
\end{deluxetable*}

\subsection{\texttt{ExoTiC-JEDI} Data Reduction} \label{sec:2.2}

The second reduction and data analysis was performed using \texttt{ExoTiC-JEDI} \citep{Alderson2022_jedi}, treating NRS1 and NRS2 separately. We followed the procedures outlined in previous \texttt{ExoTiC-JEDI} analyses (\citealt{Alderson_2023}; \citetalias{May_2023}; \citealt{Alderson_2024}). For each relevant step in the reduction, we tried a variety of parameter values, determining that the default values resulted in the lowest out-of-transit scatter in the final broadband light curves. 

Following the standard set-up of \texttt{ExoTiC-JEDI}, we began with Stages 1 and 2, performing the \texttt{jwst} calibration pipeline (v.1.14.0, context map jwst\_1242.pmap) linearity, dark current, and saturation corrections, jump detection and ramp fitting, as well as a custom bias subtraction and group level 1/$f$ noise destriping. We also obtained the 2D wavelength maps to produce wavelength solutions. To go from 2D images to stellar spectra time series, we proceeded with Stage 3, correcting for any pixels with data quality flags do not use, saturated, dead, hot, low quantum efficiency or no gain value, as well as any other spatial ($>$6$\sigma$) and temporal ($>$20$\sigma$) outliers using the default window sizes \citep[see e.g.,][]{Alderson_2024}. Any additional remaining 1/$f$ noise and background were removed by subtracting the median count value of the unilluminated pixels in each column. The location of the spectral trace was identified and an aperture five times the FWHM was used for both NRS1 and NRS2, approximately 8 pixels wide from edge to edge. The extracted stellar spectra were also cross-correlated to produce $x$- and $y$-positional shifts to be used to detrend the resulting light curves. 

We fit broadband light curves and spectroscopic light curves at a variety of resolving power for both detectors, with the broadband light curves spanning 2.814--3.717\micron\, for NRS1 and 3.824--5.111\micron\, for NRS2. For both the broadband light and spectroscopic light curves, we fitted for $R_{P}/R_{\star}$ and $t_0$, holding $a/R_{\star}$, $P$, the eccentricity ($e$) and the system inclination ($i$) fixed to values presented in Valenti et al., in prep. The stellar limb darkening coefficients were held fixed to values calculated using \texttt{ExoTiC-LD} \citep{Grant2024} based on stellar [Fe/H]=0.05, T$_\star$=5100.0, log(g)=4.416 using the \citet{kipping2013uninfomativePriorsQuad} parameterization of the square-root law and the \citet{magic2015stagger} 3D stellar model grid. 

We simultaneously fit the light curves with a \texttt{batman} \citep{Kreidberg_2015} transit model and systematic model ($S(\lambda)$) using a least squares optimizer. Our systematic model took the form 
$$
S(\lambda) = s_0 + (s_1\times t) + (s_2 \times x_s|y_s|),
$$
where $s_0$, $s_1$, $s_2$ are coefficient terms, $x_s$ is the $x$-positional shift of the spectral trace, $|y_s|$ is the absolute magnitude of the $y$-positional shift of the spectral trace and $t$ is the time. We removed any outliers greater than $2.5\sigma$ from the median of the residuals, refitting the light curves until no such points remained, and removed the first 15 integrations ($\sim$ 11 minutes) due to a settling ramp at the beginning of the observation. We additionally rescaled the light curves using the beta value \citep{Pont2006}, measured from the white and red noise values calculated by \texttt{ExoTiC-JEDI} to account for any remaining red noise. The fitting results for the \texttt{ExoTiC-JEDI} reduction are shown in Table \ref{table:fit_jedi}

\begin{deluxetable*}{lCC}
\label{table:fit_jedi}
\tablecaption{Limb darkening coefficients and light curve fit values for NRS1 and NRS2 broadband light curves from the \texttt{ExoTiC-JEDI} data reduction. }
\tablehead{ 
    \colhead{Parameters} & \colhead{NRS1} & \colhead{NRS2} 
}
\startdata
\hline
Limb darkening coefficient, $q_1$ (fixed) & 0.071961 & 0.047242 \\
Limb darkening coefficient, $q_2$ (fixed) & 0.177174 & 0.196289 \\
\hline
Planet-to-star radius ratio, $R_{P}/R_{\star}$ & 0.07192$\pm$0.00008 &0.07256$\pm$0.00009 \\ 
Mid-transit time (BJD$_\mathrm{TDB}$), $t_0$ & 2460110.810717$\pm$0.000036 & 2460110.810684$\pm$0.000042 \\
Coefficient term, $s_0$ & 35.37 $\pm$ 5.08 & -10.44 $\pm$ 5.95 \\
Coefficient term, $s_1$ & 0.0000089 $\pm$ 0.0000053 & 0.0000043 $\pm$ 0.0000067 \\
Coefficient term, $s_2$ & -0.00057 $\pm$ 0.00008 & 0.00019 $\pm$ 0.00009 \\
\hline
\enddata
\end{deluxetable*}

\begin{figure*}
    \centering
    \includegraphics[width=\textwidth]{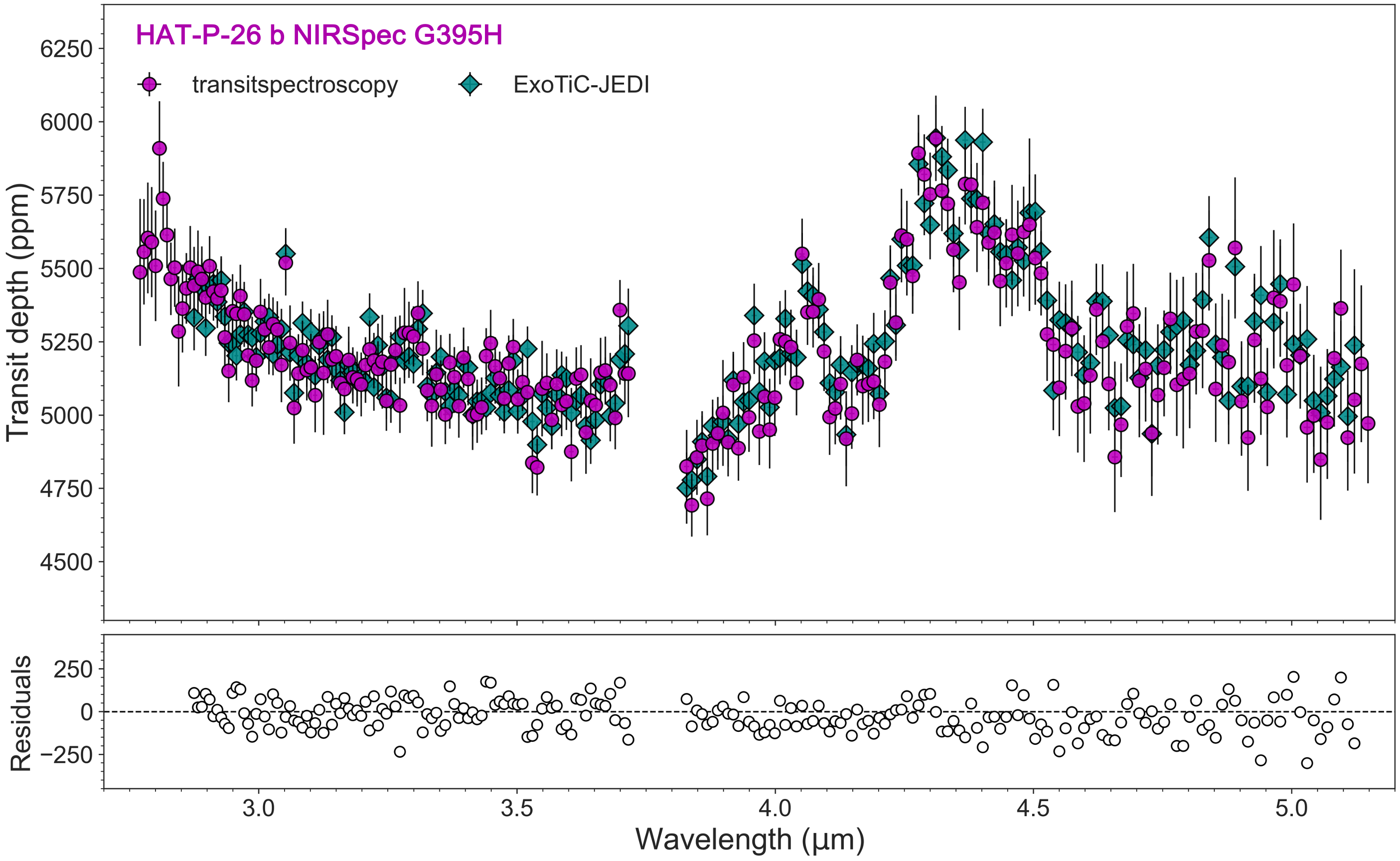}
    \caption{Two independent reductions of the HAT-P-26\,b NIRSpec G395H transit observations. Top: Transmission spectra from the \texttt{transitspectroscopy} (magenta) and \texttt{ExoTiC-JEDI} (dark cyan) pipelines, with 1$\sigma$ uncertainties binned to a resolving power of R$\sim$400. Bottom: Residuals (in ppm) between the \texttt{transitspectroscopy} and \texttt{ExoTiC-JEDI} transmission spectra. The two data reduction agree at 1$\sigma$ and show a similar shape. The \texttt{transitspectroscopy} spectrum exhibits larger dispersion and uncertainties, due to the use of Gaussian Process detrending in the light curve fits.} 
    \label{fig:data_reductions}
\end{figure*}

Each extracted transmission spectrum is binned to a resolving power of R$\sim$400 for direct comparison between reductions (Figure,\ref{fig:data_reductions}); lower-resolution spectra at R$\sim$100 are discussed later.
Residuals between the \texttt{transitspectroscopy} and \texttt{ExoTiC-JEDI} transmission spectra are shown. The two reductions agree within 1$\sigma$ and display similar shapes and amplitudes of spectral features. Particularly we can already distinguish from both reductions the SO$_2$ and CO$_2$ spectral features centred respectively at 4.05 and 4.3$\upmu$m. The \texttt{transitspectroscopy} reduction shows larger dispersion and uncertainties, as expected due to the use of Gaussian Process detrending and the fit of limb-darkening coefficients, which are fixed in the \texttt{ExoTiC-JEDI} reduction. Despite these different approaches in the light curve fitting, both methods produce robust and consistent results.

\section{Atmospheric Models}\label{sec:interpretation}
We analyzed one transit of HAT-P-26\,b obtained with NIRSpec G395H and produced two transmission spectra using independent data reduction pipelines: \texttt{transitspectroscopy} and \texttt{ExoTiC-JEDI}. These pipelines adopt different approaches to systematic correction and light curve modeling, resulting in slightly different noise properties and uncertainty estimates. To assess the robustness of our atmospheric inferences, of our atmospheric inferences, we analyze both spectra using complementary modeling frameworks: the free-chemistry retrieval codes \texttt{TauREx}\citep{Al_Refaie_2021} and \texttt{POSEIDON} \citep{MacDonald_2017, MacDonald_2023} and the self-consistent forward model grid \texttt{PICASO}\citep{Batalha_2019,Mukherjee_2023} which also includes photochemistry calculations with \texttt{Photochem} package \citep[v0.6.5][]{Wogan_2024,Wogan_2025}.

\subsection{Free atmospheric retrieval setups} \label{sec:retrievals_taurex}
\subsubsection{\texttt{TauREx} retrieval setup }
We perform a free-chemistry atmospheric retrieval using \texttt{TauREx} \citep{Al_Refaie_2021}\footnote{\url{https://taurex3.readthedocs.io/en/latest/}} on the spectra extracted with \texttt{transitspectroscopy} and \texttt{ExoTiC-JEDI}, binned to a resolving power of R$\sim$100 for computational efficiency. We also tested higher-resolution versions of the spectra and found that they yielded consistent constraints. Molecular line lists and continuum data are provided by ExoMol \citep{Tennyson_2016, Chubb_2021, tennyson_2024}, HITEMP \citep{Tennyson_2018, Rothman_2010}, and HITRAN 2020 \citep{GORDON_2022}. We include H$_2$O \citep{Polyansky_2018}, CO \citep{li_2015}, CO$_2$ \citep{Rothman_2010}, CH$_4$ \citep{Yurchenko_2017}, NH$_3$ \citep{Yurchenko_2011}, H$_2$S \citep{Azzam_2016}, HCN \citep{Barber_2014}, C$_2$H$_2$ \citep{Chubb_2020}, C$_2$H$_4$ \citep{Mant_2018}, and SO$_2$ \citep{Underwood_2016}, with their abundances ($\log(\text{X}_{\rm VMR})$) fitted using a log-uniform prior between $-12$ and $-0.3$. H$_2$ and He fill the remainder of the atmosphere and are set to a ratio of 0.17. Rayleigh scattering and H$_2$-H$_2$, H$_2$-He collision-induced absorption are also included. The temperature-pressure profile is isothermal, with the temperature fitted uniformly between 500 and 1500\,K. Cloud coverage is modeled as grey clouds with top pressures between $10^{-3}$ and $10^7$ Pa. Stellar parameters are fixed (radius: 0.87 R$_{\odot}$, T$_{\rm star}$: 5079\,K), and the planetary radius is fitted between 0.3 and 0.9 R$_{\rm Jup}$. The atmosphere is simulated with 100 atmospheric layers spanning pressures in log scale from $10^{-3}$ to $10^7$ Pa (\(10^{-8}\) to \(100\) bar). We use the \texttt{Multinest} algorithm \citep{Feroz_2009, Buchner_2014} as the optimizer an evidence tolerance of 0.5, and 1500 live points for parameter estimation. 

\subsubsection{\texttt{POSEIDON} retrieval setup }
We perform free-chemistry retrievals using \texttt{POSEIDON} \citep{MacDonald_2017, MacDonald_2023}, adopting a similar modeling setup and prior distributions, and extend the analysis to three different configurations described below. We also use similar convergence criteria and number of live points. In \texttt{POSEIDON}, the retrieved planetary radius corresponds to the radius at 10\,bar, whereas in \texttt{TauREx}, it represents the planetary radius at the bottom of the atmosphere. We use an isothermal temperature profile, as described above, but also test a more complex temperature-pressure structure using the \citet{Madhusudhan_2009} parametrization (\texttt{POSEIDON$_{\rm TP}$}). In this case, we fit for the coefficients \(\alpha_1\) and \(\alpha_2\) within the range 0.3 to 2.0, the temperature at 1\,mbar between 300 and 3000\,K, and the pressure boundaries (in logarithmic space) from \(10^{-3}\) to \(10^7\) Pa (\(10^{-8}\) to \(100\) bar). We test for an offset between the two detectors by anchoring the transit depth values from NRS1 and allowing the NRS2 values to vary with an overall offset (\texttt{POSEIDON$_{\rm offset}$}). Finally, we address the potential stellar contamination in the transmission spectrum, also known as the transit light source (TLS) effect. Stellar heterogeneities such as facul\ae{} or starspots can affect the observed spectral features, as the stellar intensity illuminating the planet differs from the baseline stellar intensity \citep{Rackham_2018}. While this effect has been identified for M dwarfs, it could also influence the analysis of transiting exoplanets around K dwarfs — such as HAT-P-26 \citep[e.g.,][]{Sanchis_Ojeda_2011, Chadney_2017, Bruno_2019}. In particular, cooler starspots may contain H$_2$O, potentially biasing the retrieved water vapour abundance in the planet's atmosphere. Following the parametrization of stellar contamination from (\cite{Pinhas_2018, rathcke_2021} \citetalias{Moran_2023}), we retrieve the stellar heterogeneity temperature, T$_{\rm het}$, using a uniform prior between 3500\,K and 6094\,K; 
the heterogeneity coverage fraction, f$_{\rm het}$, with a uniform prior between 0 and 0.5; and the stellar photosphere temperature, T$_{\rm phot}$ with a normal prior centred on T$_{\rm star}$ and a standard deviation of 100\,K. The stellar contamination factor is calculated by interpolating PHOENIX stellar grid models \citep{Allard_2012} using the \texttt{pysynphot} package \citep{lim_2013}. This retrieval, which also includes an offset, is noted as \texttt{POSEIDON$_{\rm TLS}$}.  \\

We performed atmospheric retrievals on both the \texttt{transitspectroscopy} and \texttt{ExoTiC-JEDI} transmission spectra using a range of retrieval frameworks, including \texttt{TauREx}, \texttt{POSEIDON} (with and without offsets), and \texttt{POSEIDON$_{\rm TLS}$}. As both reductions yield consistent results, we elect to present the retrievals based on the \texttt{transitspectroscopy} spectrum in the main text.
\begin{figure*}
    \centering
    \includegraphics[width=\textwidth]{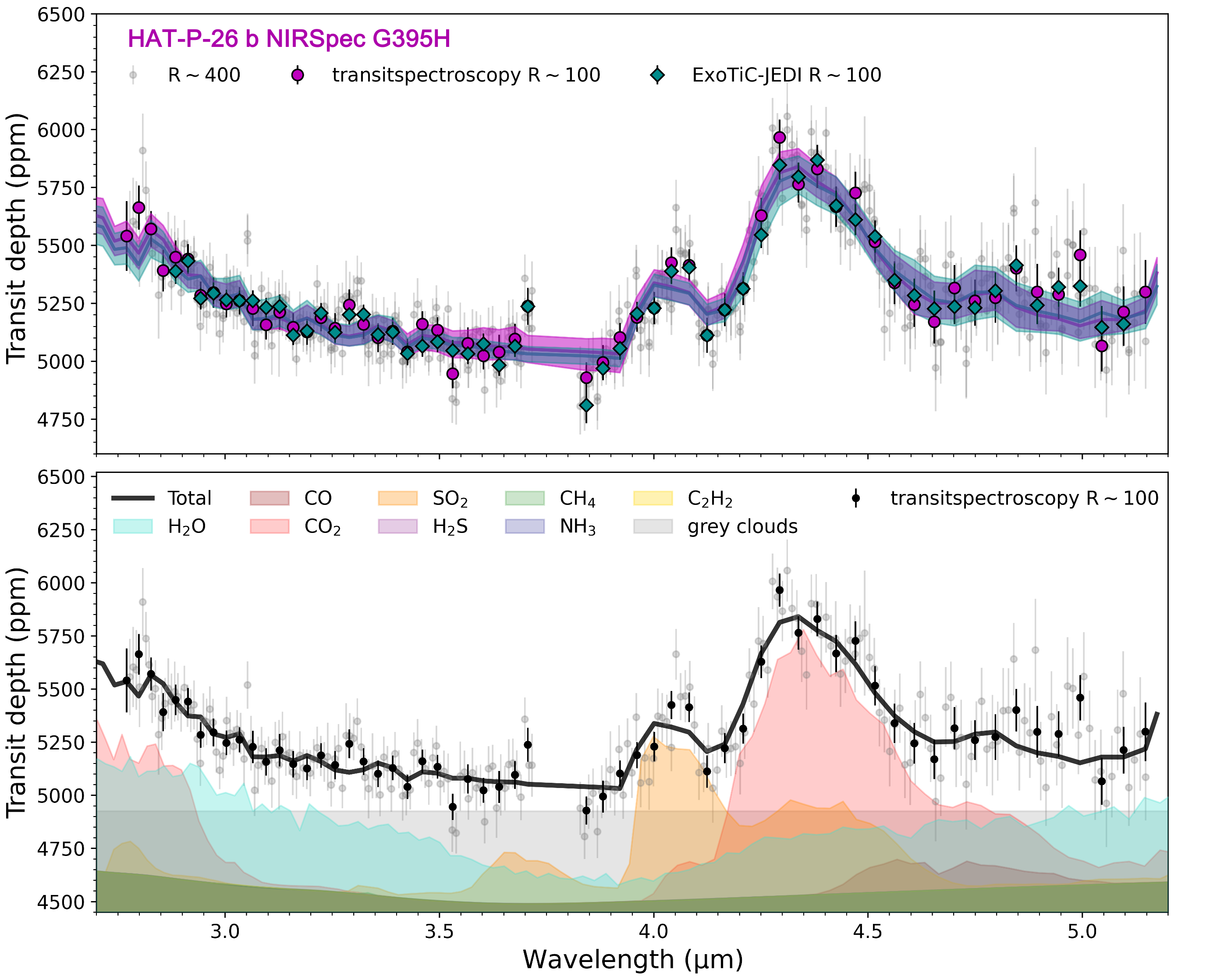}
    \caption{Top: Best-fit atmospheric retrievals and corresponding 1$\sigma$ uncertainty ranges from \texttt{POSEIDON$_{\rm TLS}$}, applied to the transmission spectra extracted using the \texttt{transitspectroscopy} (magenta) and \texttt{ExoTiC-JEDI} (dark cyan) pipelines, both binned to a resolving power of R$\sim$100. Bottom: Molecular (various colours) and grey-clouds contributions (grey) to the best-fit transmission spectrum. The black line represents the best-fit model to the \texttt{transitspectroscopy} reduction (black data points).}
    \label{fig:poseidon_retrievals}
\end{figure*}
\subsection{Retrieval results} \label{sec:retrievals_results}

\subsubsection{A note on molecular detection}
In the following discussion, we emphasize that our primary metric for evaluating molecular detections is the log Bayes factor --$\ln \mathcal{B}$. While it is common practice to translate Bayes factors into “sigma” confidence levels for ease of comparison, there is an active debate in the literature regarding how such conversions should be performed. While we usually use the \citet{Benneke_2013} formalism that follows the \citet{Trotta_2008} and \citet{Sellke_2001}  optimistic mapping, more recent discussions --e.g \citet{kipping2025} highlighted the limitations of this approach and suggested more conservative alternatives. Depending on the chosen prescription, the same Bayes factor can correspond to a wider range of “sigma” equivalents. Throughout this work we report log Bayes factors directly, report the equivalent sigma significance computed with the \citet{Sellke_2001} method, and discuss the reliability of these detections. 

\begin{deluxetable}{lC|C}[htpb]
\label{table:retrieval_best_fit}
\tablecaption{Free retrieval best-fit atmospheric and stellar properties.}
\tablehead{ 
    \colhead{Parameters}& \colhead{Priors} & \colhead{Values}  
}
\startdata
\hline
R$_{\rm P}$(R$_{\rm Jup}$) & $\mathcal{U}(0.3, 0.9)$  & 0.55$\pm$0.01  \\
T(K) & $\mathcal{U}(500, 1500)$  & 667$^{+64}_{-58}$ \\
log(H$_2$O) & $\mathcal{U}(-12, -0.3)$  &$-2.07^{+0.39}_{-0.59}$ \\
log(CO) & $\mathcal{U}(-12, -0.3)$ &  $-5.33^{+2.40}_{-4.12}$ \\
log(CO$_2$) & $\mathcal{U}(-12, -0.3)$ &$-2.91^{+0.56}_{-0.71}$  \\
log(CH$_4$) & $\mathcal{U}(-12, -0.3)$ & $<-5.49^{\Uparrow}$\\
log(SO$_2$)& $\mathcal{U}(-12, -0.3)$ &$-4.40^{+0.35}_{-0.40}$ \\
log(NH$_3$)  & $\mathcal{U}(-12, -0.3)$ & $<-4.20^{\Uparrow}$ \\ 
log(C$_2$H$_2$)  & $\mathcal{U}(-12, -0.3)$ & $<-5.33^{\Uparrow}$ \\ 
log(C$_2$H$_4$) & $\mathcal{U}(-12, -0.3)$ & $<-4.66^{\Uparrow}$ \\ 
log(H$_2$S)  & $\mathcal{U}(-12, -0.3)$ &  $-7.14^{+3.18}_{-3.08}$ \\ 
log(HCN) & $\mathcal{U}(-12, -0.3)$ & $<-4.70^{\Uparrow}$  \\ 
log(P$_{\rm clouds}$) (Pa) & $\mathcal{U}(-3, 7)$ & $2.36^{+0.60}_{-0.42}$  \\
Offset$_{\rm nrs2}$ (ppm) & $\mathcal{U}(-1000, 1000)$ & $-114^{+35}_{-35}$  \\
T$_{\rm het}$(K) & $\mathcal{U}(3500, 6094)$ & $4552^{+436}_{-289}$\\
f$_{\rm het}$ & $\mathcal{U}(0, 0.5)$ & $0.26^{+0.15}_{-0.16}$\\
T$_{\rm phot}$(K) & $\mathcal{N}(5079, 100)$ &$5104^{+89}_{-103}$ \\ \hline
MMW (amu) & derived & $2.55^{+0.32}_{-0.20}$ \\
log(H$_2$O/CO$_2$) & derived & $ 0.88^{+0.39}_{-0.49}$ \\
log(H$_2$O/CH$_4$) & derived & $<10.2^{\Uparrow}$  \\
log(CO$_2$/CH$_4$) & derived & $<9.6^{\Uparrow}$  \\
O/H ($\times$ solar) & derived  &$16.6^{+19.4}_{-11.8}$ \\
C/H ($\times$ solar) & derived & $3.8^{+8.7}_{-2.9}$  \\
S/H ($\times$ solar) & derived  &$2.6^{+5.9}_{-1.6}$ \\
C/S ($\times$ solar)  & derived & $1.29^{+2.87}_{-0.97}$ \\
O/S ($\times$ solar) & derived & $5.39^{+6.62}_{-3.61}$ \\ \hline
M/H ($\times$ solar) & derived  &$11.4^{+13.3}_{-8.1}$ \\
C/O (absolute) & derived & $0.14^{+0.21}_{-0.08}$ \\
\hline
ln$\mathcal{Z}$  & - & 459.41  \\
$\chi^2$  & - &  61.74 \\
DoF & - & 43  \\
\hline
\enddata
\tablecomments{Retrieved parameters are reported as median values with 1$\sigma$ uncertainties, or as 3$\sigma$ upper limits ($\Uparrow$). The detection significance is expressed as an x-$\sigma$ value, derived from an ablation test, i.e, removing the molecule from the model and comparing the resulting Bayesian evidence \citep{Trotta_2008, Benneke_2012}
}
\end{deluxetable}
We summarize the best-fit atmospheric properties—i.e., the retrieved and derived values from the fit with the highest $\ln\mathcal{Z}$—in Table\,\ref{table:retrieval_best_fit}. This is based on the \texttt{transitspectroscopy} spectrum retrieval using \texttt{POSEIDON$_{\rm TLS}$} setup. The full retrieval analysis for both reductions and frameworks is presented in Appendix Table~\ref{table:retrieval_master}. A comparison of the posterior distributions from the four retrieval setups is shown in Appendix Figure~\ref{fig:_retrievals_post_master} for the \texttt{transitspectroscopy} spectrum. Figure\,\ref{fig:poseidon_retrievals} shows the best-fit atmospheric model to the HAT-P-26\,b JWST NIRSpec G395H transmission spectrum for both reductions (top), along with the corresponding best-fit spectral contributions from molecular absorption and grey clouds (bottom).

\subsubsection{Best-fit results description}
Even though the model including TLS has the highest $\ln\mathcal{Z}$ (459.41), there is no statistical evidence for stellar contamination in the NIRSpec spectrum of HAT-P-26\,b, given the negligible change in log Bayes factor. When stellar contamination parameters are excluded, the model still achieves a $\ln\mathcal{Z}$ of 459.13, yielding a log Bayes factor less than 1 — indicating no statistical preference between the models with and without stellar contamination. The fraction of heterogeneities ($f_{\rm het}$) remains unconstrained. The same is true for the detector offset parameter; including an offset slightly improves the $\ln\mathcal{Z}$ (by less than 3). We note changes in the molecular abundances of the main retrieved species with or without the offset, which in turn impacts the mean molecular weight—though still within one sigma (see Table\,\ref{table:retrieval_master}). The best-fit mean molecular weight is $2.55^{+0.32}_{-0.20}$ g/mol, consistent with Neptune’s atmosphere (2.53–2.69 g/mol) and with the value inferred from HST observations in the NIR ($2.58^{+0.14}_{-0.11}$ g/mol; \citealt{MacDonald_2019}). We note that the retrieved mean molecular weight from JWST data is somewhat less constrained than the value previously inferred from HST, despite the increased precision; while we do not have a definitive explanation, this may reflect differences in the retrieval framework, such as the inclusion of additional free parameters related to TLS and the offset between instruments.

\subsubsection{Molecular absorptions}
The mean molecular weight is driven by the retrieved molecular abundances, which are linked to absorption features detected in the NIRSpec spectrum. In particular, we find moderate evidence for H$_2$O at greater than 3$\sigma$ confidence (Bayes factor $\ln \mathcal{B}=4.06$), and very strong evidence for CO$_2$ ($\ln \mathcal{B}=85.64$) and SO$_2$ ($\ln \mathcal{B}=13.46$) at greater than 5$\sigma$ confidence. We obtain these significance levels using the \citet{Sellke_2001} method and acknowledge that they could be lower with other approaches. However, we emphasize that the CO$_2$ and SO$_2$ detections are unambiguous in the spectrum, while the H$_2$O signal is more challenging but has already been detected with HST and will be confirmed strongly using SOSS (MacDonald et al., in prep.). We also find marginal evidence ($<2\sigma$) for the presence of H$_2$S and CO, with $\ln \mathcal{B}<0.5$ in both cases. While these low detection significances do not constitute formal constraints, the retrieval posteriors exhibit moderately peaked distributions for these molecules, suggesting that the data may contain weak preference for their presence depending on the chosen model framework. 
The  posterior distributions on the abundance of log(H$_2$S) is peaked at roughly -3.5 with a tail and a 3$\sigma$ upper limit of -2.67. The abundance of H$_2$S is correlated with the cloud top pressure and the presence of an offset between the instruments -- which makes it difficult to confirm (see Figure\,\ref{fig:_retrievals_post_master}). The free retrieval suggests the presence of clouds, with an opaque layer located at approximately 10$^{-2}$ bar in the atmosphere. The retrieved temperature of \(667^{+64}_{-58}\,\text{K}\) is significantly lower than the equilibrium temperature (\(\sim1000\,\text{K}\)), suggesting a cold terminator and possibly asymmetric limb conditions. Limb asymmetries in temperature and/or cloud coverage have been observed for hot and ultra-hot Jupiters using high-resolution spectroscopy \citep[e.g.,][]{Louden_2015, Ehrenreich_2020, Prinoth_2022}, but also JWST observations at lower resolutions \citep[e.g.,][]{Rustamkulov_2023, Espinoza_2024}. While modeling of asymmetric limbs has been carried out, it has so far focused exclusively on hot and ultra-hot Jupiters \citep[e.g.,][]{Fortney_2010, Line_2016b, Powell_2019, Pluriel_2022}, HAT-P-26\,b might also be subjected to inhomogeneities due to cloud condensation on the night side evaporating on the day side. A detailed investigation of potential limb asymmetries is currently underway as part of the broader transmission spectrum analysis by the JWST TST DREAMS team (Maguire et al., in prep). We also tested a more complex temperature-pressure profile using the parametrization of \citet{Madhusudhan_2009}, which yielded similar atmospheric constraints. The retrieved temperature structure is close to the isothermal profile at roughly 700\,K (see Appendix Table\,\ref{table:retrieval_master} and Figure\,\ref{fig:tp_comparison}), with no improvement in the Bayesian evidence (\(\ln\mathcal{Z} = 458.8\), compared to \(\ln\mathcal{Z} = 459.4\) from the best-fit model see also Table\,\ref{table:retrieval_best_fit}).

Water was detected in HAT-P-26\,b using HST, and carbon dioxide was predicted from equilibrium chemistry \citep{Wakeford_2017, Tsiaras_2018, MacDonald_2019}. The detection of carbon dioxide in the NIRSpec G395H data is driven by two absorption features\,: 2.7 to 3\,$\upmu$m and the strong 4.2 to 4.6\,$\upmu$m. We are able to recover the feature at the edge of the NRS1 detector wavelength range illustrated in the molecular opacity contributions seen in the second panel of Figure\,\ref{fig:poseidon_retrievals}. The significance of the H$_2$O detection is only moderate based on the log Bayes factor, likely due to degeneracies between H$_2$O abundance and stellar contamination effects like the presence of star-spots impacting the short wavelengths slope. The CO$_2$ feature at shorter wavelengths, along with a strong H$_2$O detection, should be more robustly constrained in the forthcoming SOSS analysis (MacDonald et al., in prep.).

\subsubsection{Sulfur dioxide detection}
The detection of SO$_2$ as a trace gas in hot Jupiter atmospheres was first reported in WASP-39\,b \citep{Rustamkulov_2022, Alderson_2023}, where it was found at an abundance of $\log_{10}(\mathrm{SO}_2) < -5$. Subsequent detections followed in WASP-107\,b—a Saturn-like planet—with a similar abundance \citep{Dyrek_2023, Sing_2024, Welbanks_2024}, and in the sub-Neptune GJ\,3470\,b \citep{Beatty_2024}, where $\log_{10}(\mathrm{SO}_2) < -3$. In our study, the SO$_2$ absorption features corresponds to a relatively high abundance of $\log_{10}(\mathrm{SO}_2) = -4.40^{+0.35}_{-0.40}$. The detection of SO$_2$ in the atmosphere of this prototypical Neptune-like planet is significant, as it adds to the growing list of sulfur-bearing species detected in the atmospheres of giant exoplanets around different type of stars, here a K-type star. Moreover, it extends sulfur chemistry studies to a new planetary regime, helping bridge the gap between sub-Neptunes and Jupiter- and Saturn-like planets. Given the strength of the SO$_2$ feature detected in the NIRSpec G395H spectrum, we expect that the MIRI data will provide additional constraints on the SO$_2$ abundance and associated photochemistry. A detailed analysis of the full transmission spectrum, including MIRI, will be presented in Alderson et al. (in prep).

\subsubsection{Metallicity and C/O ratio}
Using the precise abundance of the molecular species and in particular of water and carbon dioxide, we are able to infer a sub-solar C/O ratio of $0.14^{+0.21}_{-0.08}$ \citep{Lodders_2010}. The C/O ratio is computed as follows: 
\begin{equation}
    C/O = \frac{X_{CO_2} +X_{CO} + X_{CH_4}+ 2X_{C_2H_2} +2X_{C_2H_4} + X_{HCN} } {X_{H_2O} + 2X_{CO_2} + X_{CO} + 2X_{SO_2} }
\end{equation}

We also infer the atmospheric metallicity ($\frac{M/H}{M/H_{\odot}}$) and the enrichment in oxygen, carbon and sulfur in solar units. The metallicity is computed as follows using the abundances of the retrieved molecules and expressed in solar units by dividing by 8.59$\times$10$^{-4}$: 
\begin{equation}
M/H = \frac{O + C + S + N}{H}  \label{eq:metal}
\end{equation}
The metal enrichment in HAT-P-26\,b ranges from 3 to 25 $\times$ solar from the best-fit free chemistry retrieval result. A large uncertainty on the metallicity is not unexpected from the JWST NIRSpec G395H wavelength range. This result highlights the need for combined NIRISS SOSS and NIRSpec G395H observations to get a precise atmospheric metallicity as the SOSS optical to near-infrared wavelength range holds multiple water absorption bands, enabling a better constraint on the abundance of oxygen. Preliminary constraints on carbon and sulfur suggest a possible super-solar enrichment in HAT-P-26\,b’s atmosphere, though current uncertainties from free-chemistry retrieval remain too large for confirmation. We will discuss the implication of these values for planetary formation in the next section. 

\subsection{Forward Modelling with \texttt{PICASO} } \label{sec:forward_picaso}

In addition to the retrieval results, we also compute a grid of radiative convective chemical equilibrium models using the open source \texttt{PICASO} code \citep{Batalha_2019,Mukherjee_2023}. Our grid of models is self-consistent in temperature and chemistry but does not include the radiative effects of clouds. We compute our grid across four parameters: internal temperature (200\,K, 300\,K, 400\,K), heat redistribution (0.4--0.9, 6 values), metallicity (log [M/H], 0--2, 8 values), and carbon-to-oxygen ratio (0.25, 0.5,1,1.5$\times$Solar, where Solar C/O=0.458) for a total of 576 models. For our climate modelling we use the pre-weighted correlated-K tables computed by \citet{lupu_2021_7542068}. The opacity sources included in the climate calculations are: \ce{C2H2}, \ce{C2H4}, \ce{C2H6}, \ce{CH4}, CO, \ce{CO2}, CrH, Fe, FeH, \ce{H2}, H3+, \ce{H2O}, \ce{H2S}, HCN, LiCl, LiF, LiH, MgH, \ce{N2}, \ce{NH3}, OCS, \ce{PH3}, SiO, TiO, and VO, in addition to alkali metals: Li, Na, K, Rb, Cs. The high-resolution opacities used to compute these are also available on Zenodo \citep{lupu_2022_6600976}. The chemistry for these molecules were computed with the elemental abundances of \citet{Lodders2010} using the chemical equilibrium methodology in \citet{channon10} with a modified version of the NASA CEA code \citep{gordon1994computer}. With this grid of one-dimensional climate and chemistry profiles, we compute high resolution spectra from \texttt{PICASO}'s v3 R=15,000 resampled opacity database \citep{natasha_batalha_2025_14861730} with the opacity sources of H$_2$-H$_2$, H$_2$-He, H$_2$-CH$_4$, CH$_4$ \citep{Hargreaves_2020}, CO$_2$ \citep{HUANG2014reliable}, CO \citep{li_2015}, H$_2$S \citep{Azzam_2016}, H$_2$O \citep{Polyansky_2018}, SO$_2$ \citep{Underwood_2016}. We first use the pre-computed spectra to determine the highest likelihood cloud-free spectrum to the data. Specifically, we use the \texttt{ultranest} \citep{2021JOSS....6.3001B} code to derive posterior probability distributions of the grid parameters as well as an offset (in parts per million) to account for unknown reference pressure (total 5 parameters). 

When clouds are not included, the posterior distribution tends to the highest metallicity, highest C/O ratio, and lowest temperature grid point (Tint=200\,K, heat redis.=0.5, M/H=100$\times$Solar, C/O=1.5$\times$Solar). The highest likelihood model ($\ln$Z=$-375$) also has a relatively high $\chi/N$=1.3, when compared to the retrieval results. This likely suggests that the effect of clouds are needed to mute spectral features and that our cloud-free run is compensating to fit the magnitude of the \ce{CO2} feature by finding a low temperature, high metallicity, high C/O solution. Therefore, we also aim to include the effect of clouds within the \texttt{PICASO} fitting framework.

In order to incorporate the effect of clouds, we fit for the original four grid parameters and introduce an additional parameter to describe a grey extinction cloud scaling factor. Following the methodology in \citet{Rustamkulov_2023}, the cloud extinction is computed as $\tau_\mathrm{cld,i}=10^\kappa * P_i$, where $\kappa$ is a free scaling parameter and $P_i$ is the pressure at each layer. In addition, for consistency with our free-chemistry retrievals, we allow for two further parameters: a radius scaling factor ($\times$R$_p$, defined at 10-bar) and an offset between NRS1 and NRS2. The offset accounts for a detector systematic affecting the relative flux calibration of the two detectors, and is treated independently of clouds. We add the offset by adjusting the model where $>$3.7$\mu$m (region of NRS2) such that a negative offset effectively decreases the transit depth in the NRS2 region. These results provide an overall better fit to the data ($\chi/N$=1.16, $\ln$Z=$-148$), when compared to the cloud free grid posteriors ($\Delta \ln$Z=227, very strongly preferred). The retrieved offset is $-90\pm20$ppm, consistent with the offset found with the free chemistry retrieval, $-114\pm35$ppm (see Table\,\ref{table:retrieval_best_fit}). 

\begin{figure}
    \centering
    \includegraphics[width=\columnwidth]{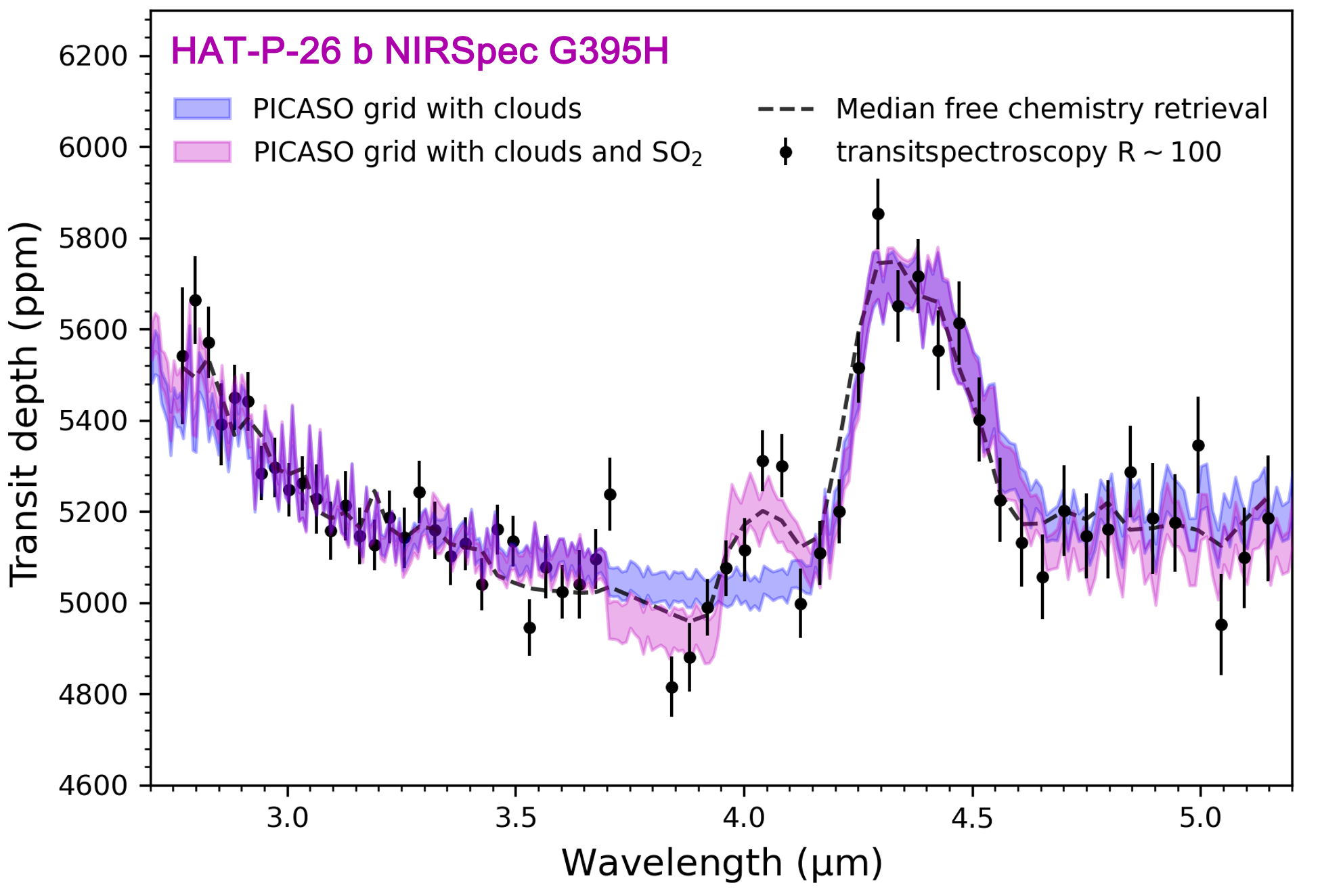}
    \caption{Atmospheric forward model fits from the \texttt{PICASO} grid with corresponding 2$\sigma$ uncertainty ranges, compared to the transmission spectrum extracted with the \texttt{transitspectroscopy} pipeline (black points), binned to a resolving power of R~$\sim$100. We present the model including SO$_2$ (magenta) and without (blue). The median best-fit model from the free-chemistry retrieval is also overplotted in dotted black line for comparison.} 
    \label{fig:picaso_modeling}
\end{figure}

Lastly, we test for the detection significance of \ce{SO2} by duplicating the scaled cloud retrieval from above but adding in a constant-with-altitude \ce{SO2} profile. The full priors and derived 1$\sigma$ confidence intervals of this result are shown in Table \ref{table:grid_fit}. The SO$_2$ model ultimately, provides an overall better fit to the data ($\chi^2/N$=0.97, $\ln$Z=$-130$), when compared to the scaled cloud posteriors ($\Delta \ln$Z=18, still strongly preferred). Besides, the inferred abundance agrees with the free retrieval method (see Tables\,\ref{table:retrieval_best_fit} and \,\ref{table:grid_fit}). All common astrophysical model parameters between the with and without SO$_2$ models are all consistent within 1$\sigma$. The instrumental  parameter to account for NRS2 offset is affected by the inclusion of \ce{SO2}. When \ce{SO2} is not included the retrieved offset is -20$\pm22$ ppm, whereas with SO$_2$ the offset increases to -90$\pm20$ ppm. Because the presence of SO$_2$ increases the transit depth at 4$\mu$m, removing the \ce{SO2} requires the offset parameter to over-compensate. The spectroscopic fit results are shown in Figure\,\ref{fig:picaso_modeling}, both with and without the inclusion of SO$_2$, while Figure\,\ref{fig:grid_free_comparison} presents the molecular abundance profiles, and the thermal structure is shown in Figure\,\ref{fig:tp_comparison}.

\subsection{Photochemical Modelling with \texttt{Photochem} } \label{sec:forward_photo}
We next test whether the derived \ce{SO2} abundance of $\sim10^{-4}$ from the  retrievals and \texttt{PICASO} grid fit  is consistent with what is expected from photochemistry. In order to do this we use the \texttt{Photochem} package \citep[v0.6.5][]{Wogan_2024,Wogan_2025}. \texttt{Photochem} solves the one-dimensional reaction-advection-diffusion equation that governs photochemistry \citep[e.g., Equation B.1 in][]{Catling_2017}, and accounts for photolysis reactions, elementary chemical reactions, and vertical transport from eddy and molecular diffusion. We use the chemical network and thermodynamics shipped with \texttt{Photochem}, except we remove all chemical species and reactions involving chlorine. The modified network considered $\sim600$ reversible reactions for $\sim100$ species composed of H, He, C, O, N, and S. 

We apply \texttt{Photochem} to the maximum likelihood model from \texttt{PICASO} scaled cloud fit with $10\times$ solar metallicity, a C/O ratio of 0.21 in absolute ($\sim$ $0.5\times$ solar), a heat redistribution factor of 0.4, and a 300\,K internal temperature. Specifically, we use the \texttt{PICASO} pressure-temperature profile, and the \texttt{PICASO} elemental abundances of H, N, O, C, S and He. Furthermore, we adopt the MUSCLES UV spectrum for HAT-P-26 \citep{Behr_2023}, and assume the eddy diffusion coefficient in Equation 2 of \citet{Tsai_2023}.


Figure \ref{fig:grid_free_comparison} shows that \texttt{Photochem} predicts a peak SO$_2$ volume mixing ratio of $10^{-4}$ at $2 \times 10^{-5}$ bar, a value compatible with $\sim 10^{-4}$ derived in the retrieval and \texttt{PICASO} analyses. By analysing the reaction rates of progress (in molecules cm$^{-3}$ s$^{-1}$) that produce and destroy SO$_2$ and its photochemical precursors near $10^{-5}$ bar, we find that SO$_2$ forms via the same pathway as described in \citet{Tsai_2023}. Through a series of six elementary reactions, H$_2$S, upwelled from the hot interior, is oxidized by H$_2$O to form SO$_2$ and H$_2$ (i.e., the net reaction is $\mathrm{H_2S} + 2\:\mathrm{H_2O} \rightarrow \mathrm{SO_2} + 3\:\mathrm{H_2}$).



\begin{deluxetable}{lC|C}[htpb]
\label{table:grid_fit}
\tablecaption{Grid-fitting best-fit atmospheric properties for cloud + SO$_2$ \texttt{PICASO} model.}
\tablehead{ 
    \colhead{Parameters}& \colhead{Priors} & \colhead{Values}  
}
\startdata
\hline
T$_{\rm int}$ (K) & $\mathcal{U}(200, 400)$  & 299$^{+69}_{-68}$ \\
Heat redistribution  & $\mathcal{U}(0.4, 0.9)$  & 0.40$^{+0.20}_{-0.01}$ \\
log M/H ($\times$ solar) & $\mathcal{U}(0, 2)$   &  1.00$^{+0.10}_{-0.10}$ \\
C/O absolute & $\mathcal{U}(0.11, 0.69)$ & 0.40$^{+0.10}_{-0.10}$ \\
log(SO$_2$)& $\mathcal{LU}(-12, -4)$ &  $-4.29^{+0.18}_{-0.26}$  \\
$\kappa$ & $\mathcal{LU}(-3,6)$ & 1.00$^{+0.10}_{-0.20}$  \\
$\times$ R$_{\rm P}$ & $\mathcal{LU}(0.7,1.3)$ & 0.76$^{+0.01}_{-0.01}$  \\
NRS2 [ppm] & $\mathcal{LU}(-600,600)$ & -90$^{+20}_{-20}$  \\
\hline
\enddata
\tablecomments{Priors are denoted as $\mathcal{U}(a,b)$ for uniform and $\mathcal{LU}(a,b)$ for log-uniform distributions over the interval [$a,b$].}
\end{deluxetable}

\section{Discussion}\label{sec:discussion}
\subsection{Disequilibrium chemistry in HAT-P-26b}

The detection of SO$_2$ in HAT-P-26\,b helps extend the regime of sulfur-based chemistry to Neptune-like planets ($\sim$20\,M$_\oplus$) and low equilibrium temperatures ($<$1000\,K). A comparable detection has been made for GJ\,3470\,b, a low-temperature sub-Neptune ($\sim$10\,M$_\oplus$, 600\,K) with JWST NIRCam transit observations \citep{Beatty_2024}. However, photochemical models predict that SO$_2$ should be more difficult to detect in lower-mass and colder planets. \citet{Tsai_2023} showed that for the same metallicity and C/O ratio as WASP-39\,b, the oxidation reactions required to produce SO from H$_2$S and OH would proceed over an order of magnitude more slowly at a temperature around 600\,K. The temperature retrieved for HAT-P-26\,b in the free retrieval is around 700\,K. However, they also show that a large abundance of H$_2$O in the atmosphere can compensate for the low temperatures via H$_2$O photolysis. 

Water photolysis is a key source of atomic hydrogen, which initiates the chemical pathway leading to SO$_2$ formation. The high water and sulfur dioxide abundances detected in GJ\,3470\,b (log(H$_2$O) = $-1.08^{+0.43}_{-0.52}$, log(SO$_2$) = $-3.57^{+0.26}_{-0.27}$) support this mechanism. \citet{Beatty_2024} showed that to produce the observed amount of SO$_2$ in GJ\,3470\,b's atmosphere requires a metallicity between 10 and 300 times solar and a C/O ratio below 0.35. This disequilibrium chemistry process could explain the molecular abundances retrieved in the atmosphere of HAT-P-26\,b (log(H$_2$O) = $-2.07^{+0.39}_{-0.59}$, log(SO$_2$) = $-4.40^{+0.35}_{-0.40}$), at a lower abundance. The retrieval findings for HAT-P-26\,b show a metallicity between 3 and 25 times solar and a sub-solar C/O ratio of 0.14, which is consistent with the formation of a substantial amount of SO$_2$ via water photolysis.

\begin{figure*} [htpb!]
    \centering
    \includegraphics[width=\columnwidth]{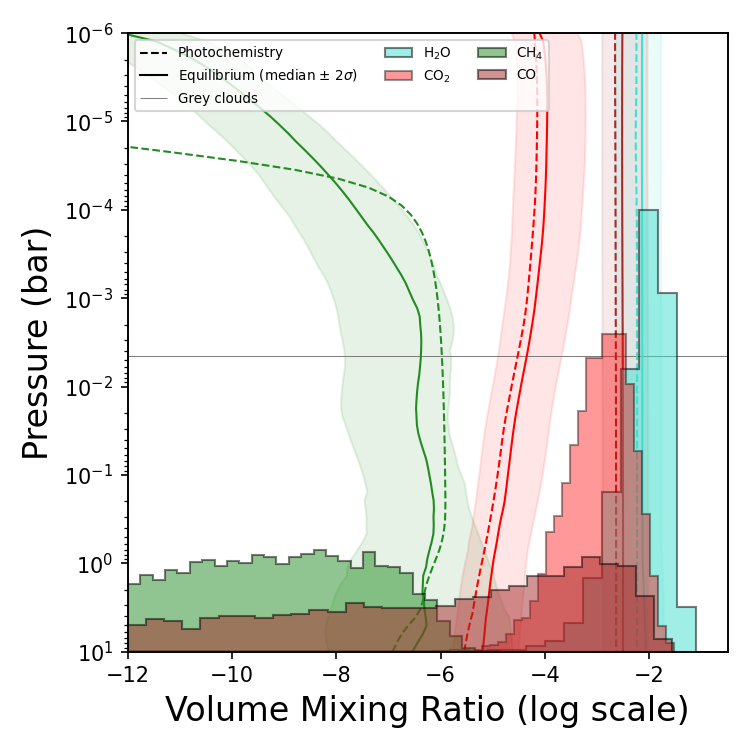}
    \includegraphics[width=\columnwidth]{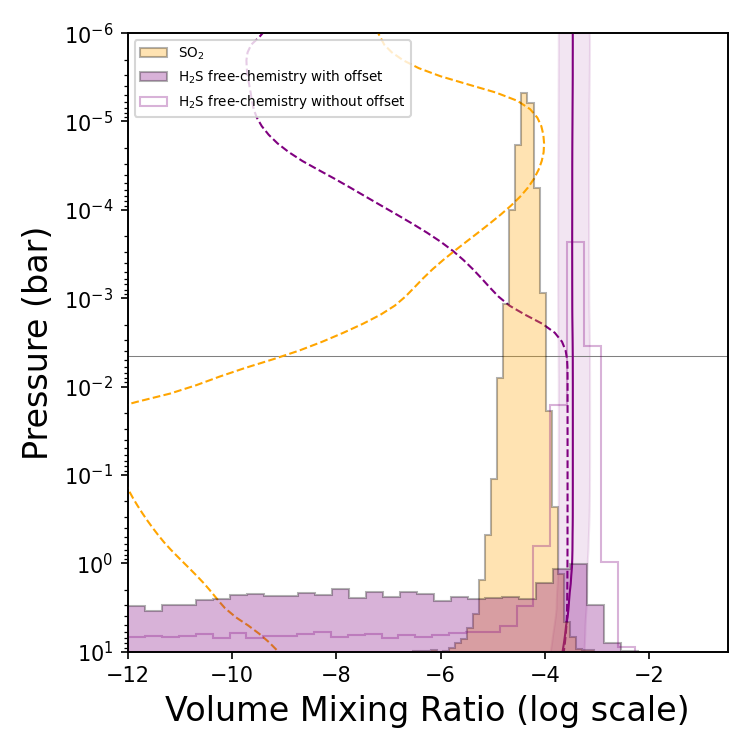}
    \caption{Comparison of results from the \texttt{PICASO} grid and free-retrieval atmospheric modeling. Colored lines show abundance profiles from the \texttt{PICASO} grid as a function of pressure: solid for chemical equilibrium, dotted for models including photochemistry. The shaded 2$\sigma$ regions represent the credible intervals of the vertical mixing ratio profiles from the grid-retrieval across the \texttt{PICASO} model grid. Posterior distributions from the free retrievals are shown as histograms; while plotted at the bottom of the panels for clarity, they represent average abundances primarily constrained between $\sim$10$^{-2}$ and 10$^{-4}$ bar. The grey horizontal line indicates the free retrieval best-fit cloud top pressure. Left: carbon-bearing species and H$_2$O. Right: sulfur-bearing species. For log(H$_2$S), we show posteriors from models with (filled) and without (unfilled) a fitted NRS1–NRS2 offset.
    }
    \label{fig:grid_free_comparison}
\end{figure*}

In Figure\,\ref{fig:grid_free_comparison}, we compare the results from the \texttt{PICASO} grid with those from the free-retrieval atmospheric modeling. The posterior distributions from the free retrievals are shown alongside the abundance profiles from the grid models as a function of pressure. Solid lines and 2$\sigma$ shaded regions represent the expected compositions under chemical equilibrium, while dotted lines show the corresponding photochemical predictions. Although the posteriors are plotted near the bottom of the figure for clarity, they represent abundances primarily constrained at pressures between $\sim$10$^{-2}$ and 10$^{-4}$ bar. Results for the carbon-bearing species and water are shown in the left panel, while sulfur-bearing species are presented in the right panel. 

The free-chemistry retrieval, even if it does not capture the vertical distribution of the molecules, broadly agree with self-consistent computations. The retrieved water abundance and the upper limits on carbon monoxide and methane are consistent with expectations from equilibrium chemistry. Similarly, the sulfur dioxide abundance agrees with predictions from photochemical models in the upper atmosphere. While we retrieve a relatively high abundance of carbon dioxide in the free-chemistry retrieval, this result remains consistent within 2$\sigma$ of the lower abundance predicted by the \texttt{PICASO} grid. This translates to a larger absolute C/O ratio fitted with \texttt{PICASO} 0.4$\pm$0.1 compared to the derived value from the free retrieval $0.14^{+0.21}_{-0.08}$. These values, though statistically consistent within 1 or 2$\sigma$, highlight degeneracies between the C/O ratio, atmospheric temperature, and cloud location. The grid model abundances are computed assuming a temperature-pressure profile hotter than the 600–700\,K isothermal structure retrieved in the free retrieval (see Figure\,\ref{fig:tp_comparison}). This mismatch in thermal structure may contribute to the discrepancies observed between the free-chemistry and grid-based abundance and cloud top pressure estimates, and could also reflect the influence of additional disequilibrium processes. 


\begin{figure*}
    \centering
    \includegraphics[width=\textwidth]{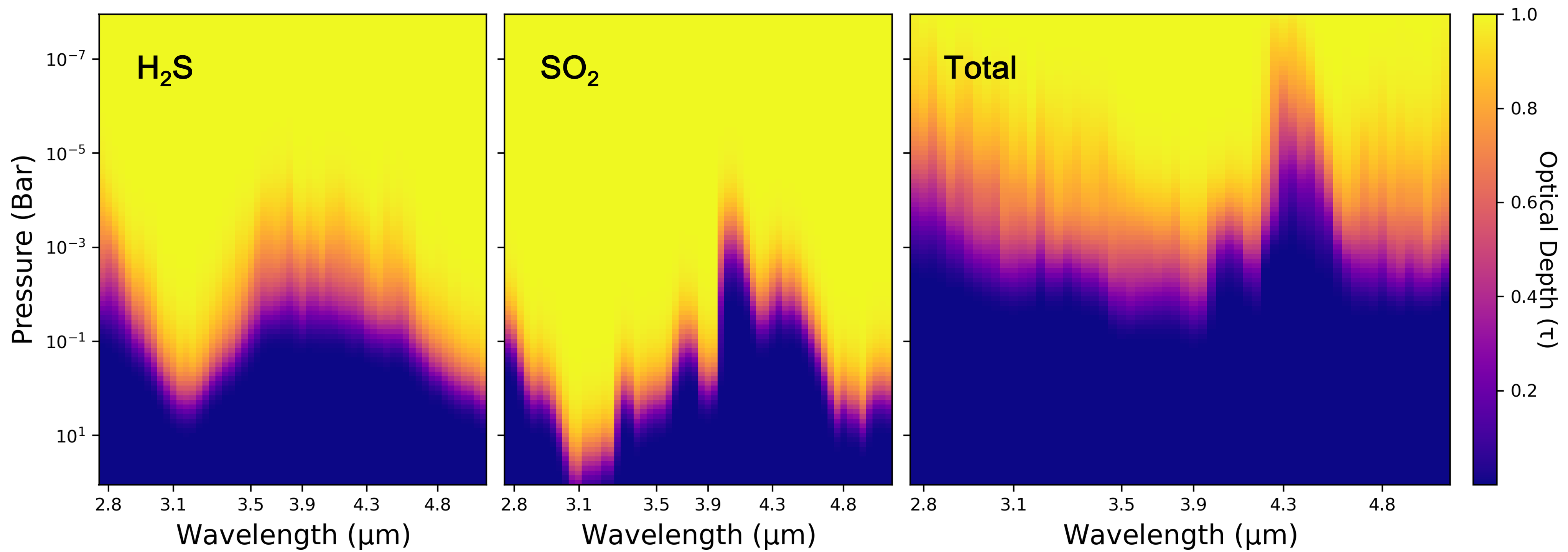}
    \caption{Opacity contribution functions for H$_2$S (left), SO$_2$ (middle), and the full atmospheric model (right) derived from the free retrieval with no offset on the \texttt{transitspectroscopy} reduction. Each panel shows the wavelength-dependent absorption contribution across atmospheric pressure levels.}
    \label{fig:contributions}
\end{figure*}

We show on the right panel of Figure\,\ref{fig:grid_free_comparison} the upper limit on H$_2$S's abundance. This panel displays the posterior distribution of the H$_2$S volume mixing ratio (VMR), both with an offset between NRS1 and NRS2 applied (filled) and without (unfilled). Without the offset, the posterior shows a slight uptick near \(\log_{10}(\mathrm{VMR}) = -3.5\). Identifying H$_2$S is inherently difficult, as its primary absorption band overlaps with that of water between 2.8 and 3.1\,$\upmu$m, while its secondary feature falls in the wavelength gap between the two detectors of NIRSpec G395H. Figure\,\ref{fig:contributions} illustrates the pressure-dependent contribution functions for H$_2$S, SO$_2$, and the total opacity, as derived from the free retrieval without applying an inter-detector offset. The total contribution is primarily dominated by H$_2$O and CO$_2$, while both H$_2$S and SO$_2$ show opacity contributions at similar pressure levels, extending up to \(10^{-5}\) bar. According to the \texttt{Photochem} models with photochemistry, H$_2$S starts to be depleted at these pressures compared to the equilibrium case. The retrieved SO$_2$ abundance aligns with these photochemical predictions. If the presence of H$_2$S is real, it could point to the influence of additional disequilibrium processes, such as vertical transport.

\subsection{Implication for HAT-P-26b's formation}
Sulfur is a valuable tracer for planetary formation, as discussed by \citet{Turrini_2021} and \citet{Crossfield_2023}. They suggest that volatile-to-refractory abundance ratios (e.g., C/S, O/S) can be used to constrain both the formation location of a planet and the formation process—whether through pebble or planetesimal accretion. Pebble accretion, which dominates beyond the evaporation line of FeS (typically $>$1–2\,AU), tends to yield supersolar C/S and O/S ratios—up to 3–20$\times$ solar because volatile elements accrete more efficiently than refractory species like sulfur. In contrast, planetesimal accretion, more common at $\lesssim1$\,AU, delivers both volatile and refractory elements more uniformly, resulting in near-solar abundance ratios. Current atmospheric constraints for HAT-P-26\,b (see Table\,\ref{table:retrieval_best_fit} suggest C/S ratios between 0.3 and 4$\times$ solar, and O/S ratios between 2 and 12$\times$ solar, potentially consistent with pebble accretion at 3–10\,AU from the host star. However, the uncertainties in these parameters remain too large to definitively rule out alternative formation pathways, and the sub-solar C/O ratio (0.15–0.35$\times$ solar) remains difficult to explain in both scenarios. We expect improved constraints from the combined analysis of all transit and eclipse observations of HAT-P-26\,b from JWST (MacDonald et al. in prep, Maguire et al. in prep). The joint SOSS, NIRSpec, and MIRI transmission spectra should yield tighter constraints on the planet's metallicity and C/O ratio (Alderson et al. in prep). One caveat is that most existing formation models have been developed for hot Jupiters; this study opens a path toward understanding the formation history of Neptune-like planets. 

The sub-solar planetary C/O ratio is broadly consistent with measurements of the stellar C/O for HAT-P-26. \citet{Kolecki_2022} report a stellar C/O of $0.89\pm0.15$, whereas \citet{polanski_2022} find C/O $= 0.41\pm0.08$, the latter being in good agreement with the planetary value of $0.41\pm 0.1$ reported in Table\,\ref{table:grid_fit}. We note, however, that this planetary C/O is derived from our grid-fitting analysis, which explores a pre-computed set of self-consistent forward models. Because the molecular abundances in the grid are constrained by the model parameters (metallicity, C/O, temperature), the formal uncertainties on C/O from the grid are inherently small. This can explain why the planetary C/O ratio appears to have a precision comparable to that of the stellar measurement. The free chemistry retrieval yields larger uncertainties on C/O, though these also depend on the retrieval setup and the specific molecules included.

\subsection{Sulfur dioxide abundance as a function of atmospheric metallicity}
The increased wavelength coverage of JWST instruments now allows for the detection of sulfur-bearing species in the atmospheres of giant exoplanets. In particular, the absorption feature of SO$_2$ at 4.05\,$\upmu$m has been identified in a small but growing number of planets, ranging from sub-Neptunes to Jupiter-like exoplanets. The robust detection of SO$_2$ ($\ln \mathcal{B} = 13.46$) in the atmosphere of HAT-P-26\,b adds to these key JWST results. The recent study by \citet{Fu_2025} compiled eight transmission spectra of giant exoplanets observed with JWST in the 3–5\,$\upmu$m range and analyzed four spectral bands that trace key molecules. Their initial population-level analysis revealed correlations between the strength of the SO$_2$ absorption feature and both planetary mass and temperature, indicating that SO$_2$ is more likely to be present in low-mass ($\lesssim$0.3\,M$\mathrm{J}$) and cooler ($\lesssim$1200\,K) targets.

\begin{figure*}
    \centering
    \includegraphics[width=\textwidth]{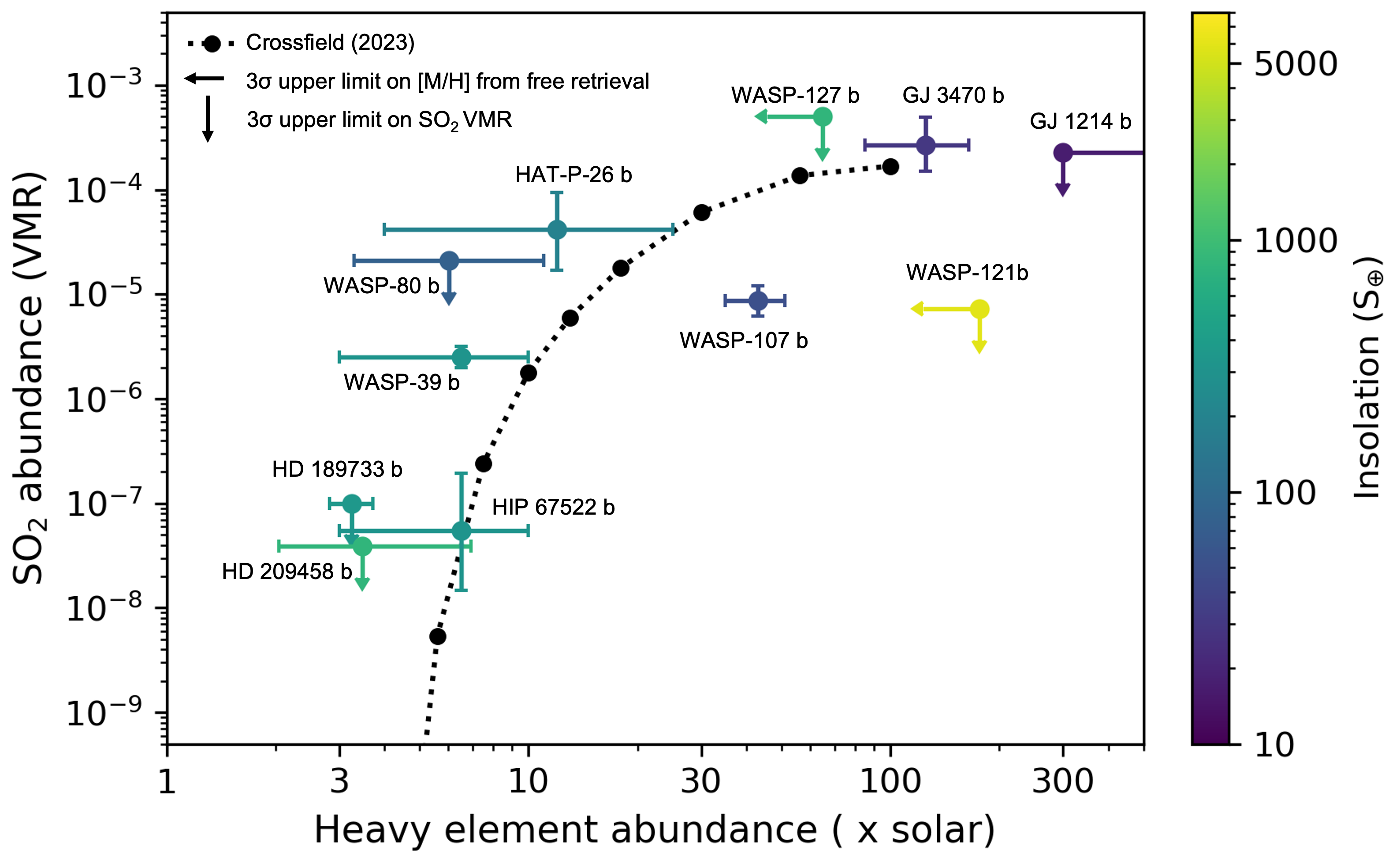}
    \caption{Predicted average SO$_2$ volume mixing ratio (VMR) at 10$^{-3}$,bar, as a function of increasing elemental abundance in log space from \citet{Crossfield_2023} (dashed black line), overplotted with JWST measurements from published transmission spectra of giant planets. SO$_2$ abundances and metallicities are taken from the literature. When SO$_2$ is not detected, 3$\sigma$ upper limits are shown. When metallicity is not provided, a 3$\sigma$ upper limit is derived from a free retrieval using \texttt{TauREx}, following a setup similar to that described in Table\,\ref{table:retrieval_master}.
    \footnotesize{References\,: GJ\,1214\,b : \citet{schlawin_2024, ohno_2025}; GJ\,3470\,b\,: \citet{Beatty_2024}; HAT-P-26\,b\,: this work; HD\,189733\,b\,: \citet{fu_2024}; HD\,209458\,b\,: \citet{Xue_2024}; HIP\,67522\,b : \citet{Thao_2024}; WASP-107\,b : \citet{Sing_2024}; WASP-121\,b\,: \citet{Fu_2025} and this work; WASP-127\,b\,: \citet{Fu_2025} and this work;  WASP-39\,b\,: \citet{Alderson_2023}; WASP-80\,b\,: \citet{bell_2023}  }
    }.\\
    
    \label{fig:fso2_met}
\end{figure*}

We adopt the same sample of spectra presented in \citet{Fu_2025} and expand the study by adding HAT-P-26\,b (this work), GJ\,1214\,b \citep{schlawin_2024, ohno_2025}, and WASP-80\,b \citep{bell_2023}. Our goal is to compare the retrieved SO$_2$ abundances to the atmospheric metallicities and test the predictions made in \citet{Crossfield_2023}. The dashed line taken from their study shows how the abundance of SO$_2$ increases as all three elements (C, O, and S) are increased in lockstep. When SO$_2$ is detected, we use the abundance constraints reported in the respective publications -- list provided in the caption of Figure\,\ref{fig:fso2_met}. The same approach is used for the metallicity, which is taken from either free retrievals or grid-model retrievals. When either of these parameters is not provided, we perform a free retrieval using a setup similar to that described in Section\,\ref{sec:retrievals_taurex}, employing the \texttt{TauREx} framework on the transmission spectra published in \citet{Fu_2025}. We then use the resulting posterior samples to report 3$\sigma$ upper limits on the SO$_2$ volume mixing ratio or the atmospheric metallicity. The metallicity is computed as in Equation\,\ref{eq:metal} using the abundances of the retrieved molecules. It is expressed in solar units by dividing by 8.59$\times$10$^{-4}$, and we provide the corresponding 3$\sigma$ upper limits. The values are overplotted in color according to their insolation and compared to the prediction in Figure\,\ref{fig:fso2_met}. The retrieved SO$_2$ abundances and metallicities for the full sample, corresponding to the data behind Figure\,\ref{fig:fso2_met}, are provided in Table\,\ref{table:fso2_met} in the Appendix.

The majority of the planets follow the predicted trend within approximately 3$\sigma$. However, the SO$_2$ volume mixing ratio (VMR) for WASP-107\,b deviates by more than 7$\sigma$, showing a lower abundance than predicted for a metallicity of 40$\times$ solar. We note that this constraint is based on a single publication using NIRSpec G395H observations \citep{Sing_2024}, despite additional constraints being available from NIRCam \citep{Welbanks_2024}, and MIRI \citep{Dyrek_2023}. A combined retrieval may therefore be necessary to better assess the metallicity and sulfur-bearing species abundances in this atmosphere. WASP-121\,b is also an outlier, which may be attributed to its extreme conditions— equilibrium temperature exceeding 2500\,K and an insolation level around 5000\,S$_{\oplus}$. These conditions place the planet in a different chemical regime, and the current free retrieval applied to its spectrum may not be adequate. A more comprehensive model that includes additional chemical species may be required to properly interpret its atmospheric composition. For planets with insolation below 2000\,S$_{\oplus}$, the current constraint on the SO$_2$ volume mixing ratio (VMR) for gas giants observed with JWST in the near-infrared follows the prediction for the complex sulfur dioxide formation pathway with a steep increase of the VMR for low metallicities and a more shallow increase above 30 $\times$ solar metallicities.

\section{Conclusions} \label{sec:conclusions}

We presented a new JWST NIRSpec G395H transit observation of the prototypical super-Neptune-sized exoplanet HAT-P-26\,b. Two independent data reduction pipelines were used, producing consistent transmission spectra despite differing choices in data processing and light curve fitting. We report the detection of water ($\ln \mathcal{B} = 4.06$), and the very strong evidence of carbon dioxide ($\ln \mathcal{B} = 85.64$), and sulfur dioxide ($\ln \mathcal{B} = 13.46$). 
It is important to note that our main molecular claims are not contingent on the precise choice of sigma conversion. The strongest detections we report, CO$_2$ and SO$_2$, are unambiguous in the data and readily identifiable by eye in the spectrum. The detection of H$_2$O, while quantitatively weaker, is fully consistent with prior HST results \citep{Wakeford_2017, Tsiaras_2018, MacDonald_2019} and will be independently confirmed in JWST/SOSS observations (Macdonald and the GTO TST DREAMS in prep.). Thus, our interpretation remains robust regardless of the exact statistical mapping adopted.

The detection of sulfur dioxide in HAT-P-26\,b—a warm Neptune—is particularly significant. SO$_2$ had only been robustly detected in a handful of hot Jupiters and a single Neptune (GJ\,3470\,b). This result demonstrates that disequilibrium photochemistry is active in the atmospheres of smaller, close-in exoplanets over a broader temperature range than previously predicted. By re-analyzing a sample of ten JWST near-infrared transmission spectra of giant exoplanets, we find that SO$_2$ abundance correlates with atmospheric metallicity, consistent with the trend predicted by \citet{Crossfield_2023}: a steep increase at low metallicities, followed by a more gradual rise at metallicities above $\sim30 \times$ solar.

Using two different Bayesian retrieval frameworks and both free-chemistry and self-consistent (photochemical) modeling, we place strong constraints on the atmospheric composition of HAT-P-26\,b. We find a metallicity between 3 and 25$\times$ solar and a sub-solar C/O ratio ranging from 0.06-0.35. The free-chemistry retrievals favor a relatively cool, isothermal temperature profile between 600 and 700\,K —significantly lower than the $\sim$1000\,K equilibrium temperature——and suggest the presence of grey clouds at pressures around 10$^{-2}$\,bar. These conclusions are robust across two frameworks, five different retrieval setups that include varying temperature parameterizations, stellar contamination models, detector offsets, and both independent data reductions. The self-consistent grid-based retrievals using a non-isothermal temperature-pressure profile return metallicities and C/O ratios broadly consistent within 1–2$\sigma$ of the free retrievals. The difference in thermal structure likely contributes to small discrepancies in retrieved molecular abundances and cloud properties and may reflect limb-averaged effects such as temperature asymmetries or inhomogeneous cloud coverage. We also retrieve a high abundance of CO$_2$ in the free-chemistry case, which, while consistent within 2$\sigma$ of the CO$_2$ prediction from the photochemical grid model, may hint at sensitivity to the thermal structure and possible contributions from disequilibrium chemistry. Finally, we identify a marginal signature of H$_2$S, which appears as a modest uptick in the posterior distribution, particularly when no offset is applied between NRS1 and NRS2. Confirming the presence of H$_2$S remains challenging due to its spectral overlap with H$_2$O, the instrumental gap near its absorption band, and its degeneracy with cloud coverage. While the constraints on other molecules are robust, the tentative H$_2$S detection at the $\sim$2$\sigma$ level is sensitive to retrieval assumptions. 
Nevertheless, its presence at similar pressure levels to SO$_2$—its expected photochemical byproduct—combined with the elevated retrieved abundance, may suggest additional disequilibrium processes, such as vertical mixing, that are not accounted for.

The forthcoming paper based on the NIRISS SOSS observation (Macdonald and the GTO TST DREAMS in prep.) and the combined analysis of the full JWST dataset—including NIRISS, NIRSpec, and MIRI— (Alderson and the GTO TST DREAMS, in prep.) will provide broader spectral coverage and new constraints on the planet's atmospheric composition. This joint analysis is expected to refine the metallicity estimate through improved measurements of oxygen-bearing species, confirm the presence of clouds, and offer deeper insights into the disequilibrium chemistry processes shaping the atmosphere of HAT-P-26\,b.

\section*{Acknowledgments}
This paper reports work carried out in the context of the GTO Science Program (Co-PIs: R.~van der Marel, M.~Perrin, N.~Lewis) of the Team (see \url{https://www.stsci.edu/~marel/jwsttelsciteam.html})
) of the JWST Telescope Scientist (M.~Mountain). Funding is provided to the team by NASA through grant 80NSSC20K0586. Based on observations with the NASA/ESA/CSA JWST, associated with program(s) GTO-1312 (PI: N. Lewis), obtained at the Space Telescope Science Institute, which is operated by AURA, Inc., under NASA contract NAS 5-03127. LA is supported by Cornell University College of Arts \& Sciences Klarman Fellowship. HRW was funded by UK Research and Innovation (UKRI) framework under the UK government’s Horizon Europe funding guarantee for an ERC Starter Grant [grant number EP/Y006313/1]. RJM is supported by NASA through the NASA Hubble Fellowship grant HST-HF2-51513.001, awarded by the Space Telescope Science Institute, which is operated by the Association of Universities for Research in Astronomy, Inc., for NASA, under contract NAS 5-26555. DRL acknowledges support from NASA under award number 80GSFC24M0006. CIC acknowledges support by NASA Headquarters through an appointment to the NASA Postdoctoral Program at the Goddard Space Flight Center, administered by ORAU through a contract with NASA. 

The JWST data presented in this paper were obtained from the Mikulski Archive for Space Telescopes (MAST) at the Space Telescope Science Institute. The specific observations analyzed can be accessed via  \dataset[DOI: 10.17909/c23k-dn78]{http://dx.doi.org/10.17909/c23k-dn78}.


\vspace{5mm}
\facilities{JWST(NIRSpec G395H)}

\software{ 
\texttt{transitspectroscopy} \citep{espinoza_2022} \\
\texttt{ExoTiC} \citep{Alderson2022_jedi} \\ 
\texttt{TauREx} \citep{Al_Refaie_2021} \\
\texttt{POSEIDON} \citep{MacDonald_2017, MacDonald_2023} \\
\texttt{PICASO} \citep{Batalha_2019, Mukherjee_2023} \\
\texttt{Photochem} \citep{Wogan_2024, Wogan_2025}}

\bibliography{manuscript}{}
\bibliographystyle{aasjournal}

\appendix  
\renewcommand\thefigure{\thesection.\arabic{figure}}    

\section{Free atmospheric retrievals}
This appendix presents additional materials from the free retrieval results from HAT-P-26\,b transmission spectrum using two Bayesian frameworks: \texttt{TauREx} and \texttt{POSEIDON}.

\begin{figure*}[htpb]
    \centering
    \includegraphics[width=\textwidth]{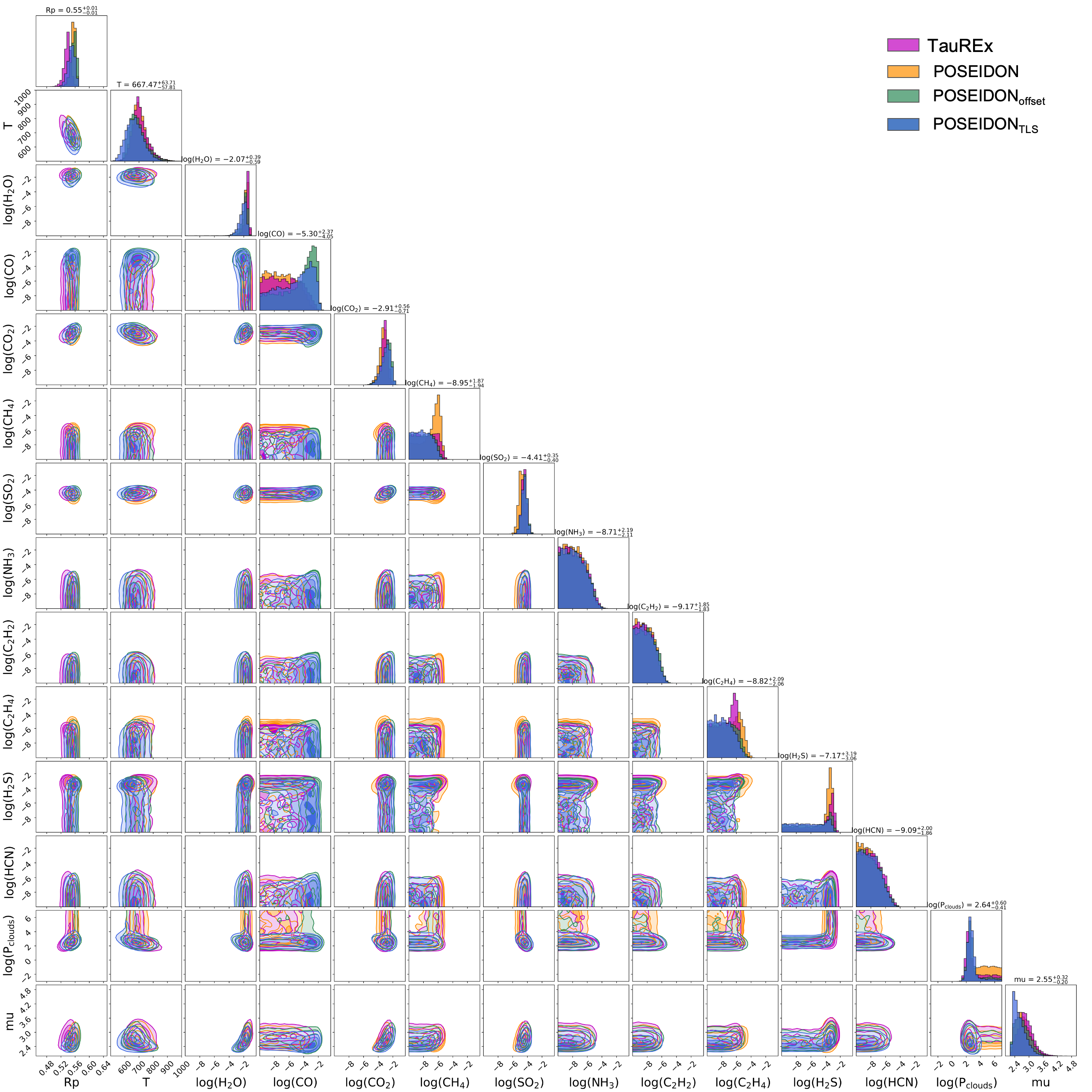}
    \caption{Posterior distributions for the free retrievals of the JWST NIRSpec G395H transmission spectrum of HAT-P-26\,b, based on the \texttt{transitspectroscopy} reduction. The different colors correspond to the various retrieval setups described in Table,\ref{table:retrieval_master}. The retrieved values shown correspond to the best-fitting parameters from the \texttt{POSEIDON$_{\rm TLS}$} setup (blue). }
    \label{fig:_retrievals_post_master}
\end{figure*}

\begin{deluxetable*}{lC|CCCC|CCC}[htpb]
\label{table:retrieval_master}
\tablecaption{Free retrieval results on HAT-P-26\,b JWST NIRSpec G395H transmission spectrum.}
\tablehead{ 
     &  & \multicolumn{4}{c}{\texttt{transitspectroscopy}} & \multicolumn{3}{c}{\texttt{ExoTiC-JEDI}} \\
  \colhead{Parameters}  &\colhead{Priors} & \colhead{\texttt{TauREx}} & \colhead{\texttt{POSEIDON}} & \colhead{\texttt{POSEIDON$_{\rm offset}$}} & \colhead{\texttt{POSEIDON$_{\rm TP}$}} & \colhead{\texttt{TauREx}} & \colhead{\texttt{POSEIDON}} & \colhead{\texttt{POSEIDON$_{\rm TLS}$}}
}
\startdata
\hline
R$_{\rm P}$(R$_{\rm Jup}$) & $\mathcal{U}(0.3, 0.9)$  & 0.54$\pm$0.01 &  0.55$\pm$0.01  &  0.56$\pm$0.01 &  0.55$\pm$0.01 & 0.55$\pm$0.01& 0.55$\pm$0.01 & 0.54$\pm$0.01\\
T(K) & $\mathcal{U}(500, 1500)$ & 701$^{+61}_{-43}$ &700$^{+59}_{-45}$ & 682$^{+69}_{-50}$  & -&696$^{+53}_{-44}$& 739$^{+98}_{-63}$&760$^{+85}_{-93}$\\
T$_{ref}$(K) & $\mathcal{U}(300, 3000)$ & - & - & - & 682$^{+69}_{-66}$& - & -& -  \\
$\alpha_1$ & $\mathcal{U}(0.3, 2.0)$  & - & - & - &1.17 $^{+0.54}_{-0.58}$ & - & -& - \\
$\alpha_2$ & $\mathcal{U}(0.3, 2.0)$ & - & - & - & 1.16 $^{+0.54}_{-0.57}$ & - & -& - \\
log(P$_{\rm 1}$) (Pa)  & $\mathcal{U}(-3, 7)$ (Pa) & - & - & - & 0.05 $^{+2.90}_{-2.00}$ & - & -& - \\
log(P$_{\rm 1}$) (Pa)  & $\mathcal{U}(-3, 7)$ (Pa) & - & - & - &  0.06 $^{+2.67}_{-2.32}$ & - & -& - \\
log(P$_{\rm 1}$) (Pa)  & $\mathcal{U}(-3, 7)$ (Pa) & - & - & - & 4.69 $^{+1.59}_{-2.26}$ & - & -& - \\
log(H$_2$O) & $\mathcal{U}(-12, -0.3)$ & $-1.55^{+0.19}_{-0.28}$ & $-1.63^{+0.24}_{-0.37}$  
&$-1.83^{+0.31}_{-0.58}$&$-2.11^{+0.42}_{-0.63}$   & $-1.30^{+0.12}_{-0.12}$ &$-1.41^{+0.17}_{-0.36}$ & $-2.06^{+0.55}_{-0.68}$ \\
log(CO) & $\mathcal{U}(-12, -0.3)$ &  $<-2.23^{\Uparrow}$ & $<-2.60^{\Uparrow}$ & $-4.39^{+1.78}_{-4.74}$ & $-5.47^{+2.52}_{-4.00}$& $<-2.14^{\Uparrow}$& $-4.71^{+1.83}_{-4.33}$& $-2.94^{+0.78}_{-3.21}$ \\
log(CO$_2$) & $\mathcal{U}(-12, -0.3)$ & $-3.08^{+0.39}_{-0.45}$ & $-3.32^{+0.44}_{-0.53}$ & $-2.78^{+0.50}_{-0.76}$ & $-2.94^{+0.61}_{-0.76}$ &$-2.38^{+0.29}_{-0.26}$ & $-2.80^{+0.38}_{-0.59}$ &$-3.31^{+0.73}_{-0.73}$  \\
log(CH$_4$) & $\mathcal{U}(-12, -0.3)$ & $<-4.59^{\Uparrow}$ & $<-5.07^{\Uparrow}$ & $<-5.35^{\Uparrow}$  & $<-5.39^{\Uparrow}$ & $<-4.99^{\Uparrow}$& $<-5.03^{\Uparrow}$& $<5.18-^{\Uparrow}$ \\
log(SO$_2$)& $\mathcal{U}(-12, -0.3)$ &  $-4.45^{+0.32}_{-0.33}$ & $-4.86^{+0.32}_{-0.34}$ & $-4.39^{+0.35}_{-0.39}$ & $-4.41^{+0.37}_{-0.39}$ &$-3.97^{+0.27}_{-0.25}$ & $-4.22^{+0.28}_{-0.27}$ &$-4.29^{+0.30}_{-0.31}$  \\
log(NH$_3$)  & $\mathcal{U}(-12, -0.3)$ & $<-4.05^{\Uparrow}$ & $<-4.43^{\Uparrow}$& $<-4.03^{\Uparrow}$& $<-4.09^{\Uparrow}$& $<-3.85^{\Uparrow}$ & $<-4.04^{\Uparrow}$ & $<-4.17^{\Uparrow}$ \\ 
log(C$_2$H$_2$)  & $\mathcal{U}(-12, -0.3)$ & $<-5.37^{\Uparrow}$ & $<-5.49^{\Uparrow}$ & $<-5.27^{\Uparrow}$& $<-5.33^{\Uparrow}$ & $<-4.72^{\Uparrow}$ & $<-4.91^{\Uparrow}$ & $<-4.97^{\Uparrow}$  \\ 
log(C$_2$H$_4$) & $\mathcal{U}(-12, -0.3)$ &  $<-5.11^{\Uparrow}$ & $<-4.06^{\Uparrow}$ & $<-4.42^{\Uparrow}$& $<-4.57^{\Uparrow}$& $<-5.06^{\Uparrow}$ & $<-3.62^{\Uparrow}$ &$<-3.90^{\Uparrow}$ \\ 
log(H$_2$S)  & $\mathcal{U}(-12, -0.3)$ & $-4.43^{+1.39}_{-5.05}$ & $-3.56^{+0.44}_{-3.30}$  &  $<-2.57^{\Uparrow}$ &  $<-2.62^{\Uparrow}$ & $-4.17^{+1.09}_{-4.99}$ & $-6.46^{+1.57}_{-3.57}$ & $<-2.85^{\Uparrow}$ \\ 
log(HCN) & $\mathcal{U}(-12, -0.3)$ & $<-4.50^{\Uparrow}$ & $<-4.92^{\Uparrow}$ & $<-4.65^{\Uparrow}$  & $<-4.61^{\Uparrow}$  &$<-4.11^{\Uparrow}$ &$<-4.30^{\Uparrow}$ &$<-4.35^{\Uparrow}$  \\ 
log(P$_{\rm clouds}$) (Pa) & $\mathcal{U}(-3, 7)$ & $2.55^{+2.41}_{-0.49}$ & $4.29^{+1.80}_{-1.72}$ & $2.62^{+1.31}_{-0.44}$ &$2.63^{+0.58}_{-0.44}$  &  $3.05^{+2.51}_{-0.93}$  & $2.56^{+2.02}_{-0.58}$ &$2.61^{+0.46}_{-0.36}$  \\
Offset$_{\rm nrs2}$ (ppm) & $\mathcal{U}(-1000, 1000)$ & - & - & $-107^{+34}_{-39}$& $-114^{+35}_{-35}$ & -  & -  & $-54^{+34}_{-34}$ \\
T$_{\rm het}$(K) & $\mathcal{U}(3500, 6094)$ & - & - & -& $4536^{+429}_{-295}$ & -  & - &$4631^{+429}_{-362}$\\
f$_{\rm het}$ & $\mathcal{U}(0, 0.5)$ &- & - & - & $0.25^{+0.15}_{-0.16}$ & -& - & $0.21^{+0.16}_{-0.13}$\\
T$_{\rm phot}$(K) & $\mathcal{N}(5079, 100)$ & - & - & - & $5101^{+90}_{-100}$ & - & - & $5111^{+86}_{-103}$ \\ \hline
MMW (amu) & derived & $2.81^{+0.29}_{-0.25}$ & $2.72^{+0.35}_{-0.27}$ &$2.69^{+0.36}_{-0.28}$ &  $2.52^{+0.30}_{-0.18}$ & $3.33^{+0.37}_{-0.25}$&  $3.02^{+0.40}_{-0.43}$& $2.60^{+0.42}_{-0.23}$ \\
\hline
ln$\mathcal{Z}$  & - & 455.70 & 457.05 & 459.13 & 458.8 &428.2 & 429.9 & 428.6 \\
$\chi^2$ & - &  70.05 & 69.25 & 62.43 & 62.10 & 70.61 & 69.21 & 65.19 \\
DoF   & - &  47 & 47 & 46 & 42 & 42  & 42 & 38 \\
\hline
\enddata
\tablecomments{Retrieved parameters are reported as median values with 1$\sigma$ uncertainties, or as 3$\sigma$ upper limits ($\Uparrow$). Parameters that are not included in the fit are indicated with a dash ($-$).
}
\end{deluxetable*}

\begin{figure}[htpb]
    \centering
    \includegraphics[width=\columnwidth]{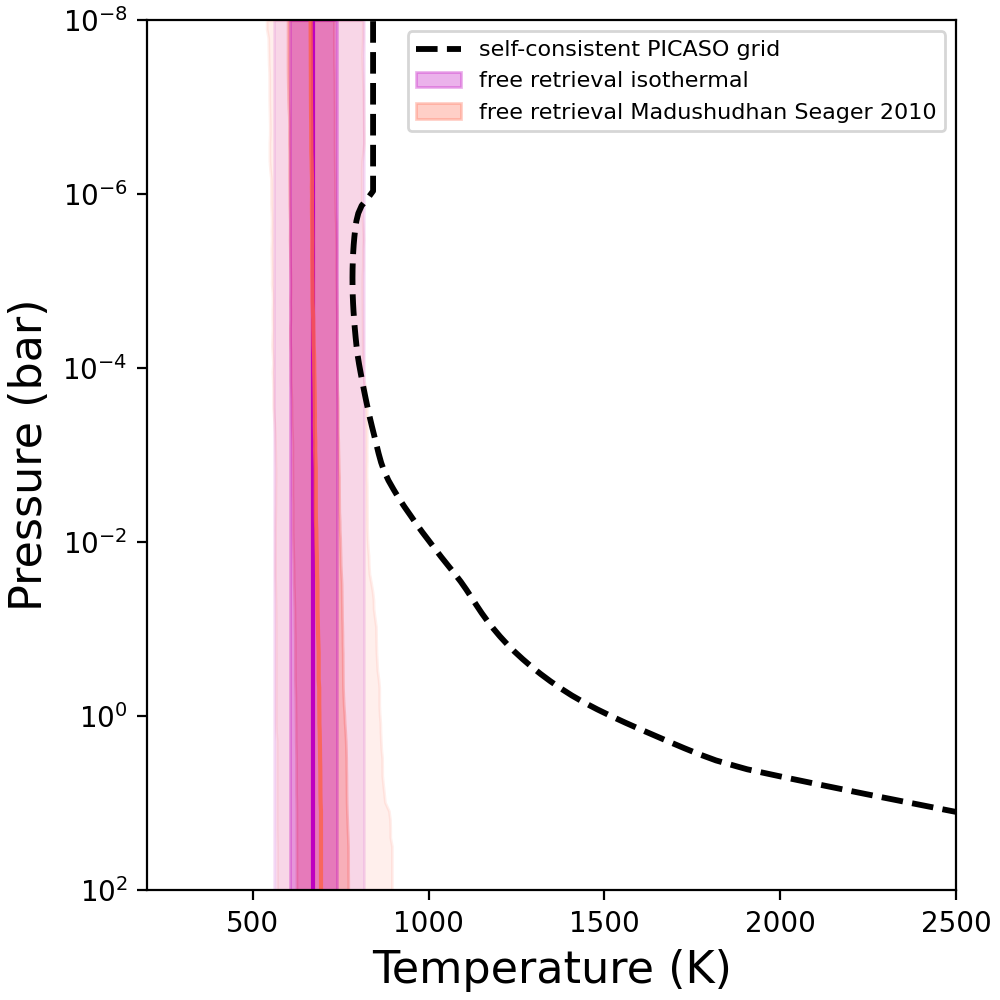}
    \caption{Comparison of the temperature–pressure structure of HAT-P-26\,b from the free retrieval analysis using an isothermal (magenta) and \citet{Madhusudhan_2009} (orange) parameterization, alongside the self-consistent computation from the \texttt{PICASO} grid (black dotted line).  }
   \label{fig:tp_comparison}
\end{figure}

\section{Sulfur dioxide abundance as a function of atmospheric metallicity}

This appendix presents the retrieved SO$_2$ abundances and metallicities for the full sample, corresponding to the data behind Figure\,\ref{fig:fso2_met}. 

\begin{deluxetable*}{lccccccccc}[htbp]
\tablecaption{Planetary and stellar parameters, together with retrieved SO$_2$ abundances and metallicities ([M/H]) for a sample of hot and warm exoplanets observed with JWST, corresponding to the data shown in Figure\,\ref{fig:fso2_met}.}
\label{table:fso2_met}
\tablehead{
\colhead{Planet} & 
\colhead{$R_\star$ [R$_\odot$]} & 
\colhead{$T_\star$ [K]} & 
\colhead{$a$ [AU]} & 
\colhead{S [S$_{\oplus}$]} & 
\colhead{log(SO$_2$)} & 
\colhead{[M/H]} & 
\colhead{$M_p$ [M$_\mathrm{J}$]} & 
\colhead{$R_p$ [R$_\mathrm{J}$]} & 
\colhead{$T_\mathrm{eq}$ [K]} 
}
\startdata
GJ\,1214\,b & 0.216 & 3101 & 0.0151 & 17.12 & $<-3.64^{\Uparrow}$ & $>300^{\Downarrow}$ & 0.0265 & 0.244 & 567 \\
GJ\,3470\,b & 0.500 & 3600 & 0.0355 & 29.85 & $-3.57^{+0.26}_{-0.25}$ & $125^{+40}_{-40}$ & 0.044 & 0.408 & 600 \\
HAT-P-26\,b & 0.860 & 5079 & 0.0479 & 192.35 & $-4.40^{+0.35}_{-0.40}$ & $11^{+13}_{-8}$ & 0.058 & 0.565 & 1000 \\
HD\,189733\,b & 0.765 & 5052 & 0.0310 & 355.91 & $<-7.00^{\Uparrow}$  & $3.24^{+0.48}_{-0.42}$ & 1.13 & 1.13 & 1209 \\
HD\,209458\,b & 1.199 & 6026 & 0.0463 & 792.01 & $<-7.41^{\Uparrow}$ & $3.47^{+3.45}_{-1.43}$ & 0.73 & 1.39 & 1459 \\
HIP\,67522\,b & 1.380 & 5675 & 0.0748 & 316.74 & $-7.26^{+0.55}_{-0.57}$ & $6.5^{+3.5}_{-3.5}$ & 0.047 & 0.891 & 1175 \\
WASP-107\,b & 0.670 & 4425 & 0.0550 & 51.04 & $-5.06^{+0.14}_{-0.15}$ & $43^{+8}_{-8}$ & 0.096 & 0.94 & 750  \\
WASP-121\,b & 1.458 & 6776 & 0.0260 & 5966.08 & $<-5.14^{\Uparrow}$ & $<176^{\Uparrow}$ & 1.157 & 1.753 & 2450  \\
WASP-127\,b & 1.347 & 5828 & 0.0484 & 801.76 & $<-3.30^{\Uparrow}$ & <$65^{\Uparrow}$ & 0.165 & 1.311 & 1427 \\
WASP-39\,b & 1.013 & 5326 & 0.0484 & 318.08 & $-5.60\pm0.10$ & $6.5^{+3.5}_{-3.5}$ & 0.28 & 1.27 & 1166 \\
WASP-80\,b & 0.586 & 4143 & 0.0344 & 76.70 & $<-4.68^{\Uparrow}$ & $6.03^{+4.99}_{-2.72}$ & 0.538 & 0.999 & 825 \\
\enddata
\tablecomments{{References\,: GJ\,1214\,b : \citet{schlawin_2024, ohno_2025}; GJ\,3470\,b\,: \citet{Beatty_2024}; HAT-P-26\,b\,: this work; HD\,189733\,b\,: \citet{fu_2024}; HD\,209458\,b\,: \citet{Xue_2024}; HIP\,67522\,b : \citet{Thao_2024}; WASP-107\,b : \citet{Sing_2024}; WASP-121\,b\,: \citet{Fu_2025} and this work; WASP-127\,b\,: \citet{Fu_2025} and this work;  WASP-39\,b\,: \citet{Alderson_2023}; WASP-80\,b\,: \citet{bell_2023}  }}
\end{deluxetable*}

\end{document}